\documentclass[conference]{IEEEtran}
\IEEEoverridecommandlockouts
\usepackage{xspace}
\usepackage{caption}
\usepackage{subcaption}
\usepackage{cite}
\usepackage{amsmath,amssymb,amsfonts}
\usepackage{enumitem}
\usepackage{graphicx}
\usepackage{textcomp}
\usepackage{xcolor}
\usepackage{xurl}
\usepackage{listings}
\usepackage{comment}
\usepackage{algorithm}
\usepackage{algpseudocode}
\usepackage{braket} 
\usepackage[hidelinks]{hyperref}
\def\BibTeX{{\rm B\kern-.05em{\sc i\kern-.025em b}\kern-.08em
    T\kern-.1667em\lower.7ex\hbox{E}\kern-.125emX}}

\usepackage{multirow}
\usepackage{listings}
\usepackage{xcolor}
\usepackage{makecell}

\definecolor{codegreen}{rgb}{0,0.6,0}
\definecolor{codegray}{rgb}{0.5,0.5,0.5}
\definecolor{codepurple}{rgb}{0.58,0,0.82}
\definecolor{backcolour}{rgb}{0.95,0.95,0.92}

\newcommand{\todo}[1]{\textcolor{red}{TODO: #1}}

\usepackage{AD/ipdps26repro}

\lstdefinestyle{mystyle}{
    commentstyle=\color{codegreen},
    keywordstyle=\color{magenta},
    numberstyle=\tiny\color{codegray},
    stringstyle=\color{codepurple},
    basicstyle=\ttfamily\scriptsize,
    breakatwhitespace=false,         
    breaklines=true,                 
    captionpos=t,                    
    keepspaces=true,                 
    numbers=left,                    
    numbersep=5pt,                  
    showspaces=false,                
    showstringspaces=false,
    showtabs=false,                  
    tabsize=1,
    frame=tb
}

\usepackage{makecell}
\usepackage{booktabs}
\begin{document}

\title{High-performance Vector-length Agnostic Quantum Circuit Simulations on ARM Processors
}
\author{\IEEEauthorblockN{Ruimin Shi}
\IEEEauthorblockA{\textit{KTH Royal Institute of Technology}\\
Stockholm, Sweden \\ ruimins@kth.se\\
}
\and
\IEEEauthorblockN{Gabin Schieffer}
\IEEEauthorblockA{\textit{KTH Royal Institute of Technology}\\
Stockholm, Sweden \\ gabins@kth.se\\
}
\and 
\IEEEauthorblockN{Pei-Hung Lin}
\IEEEauthorblockA{\textit{Lawrence Livermore National Laboratory}\\
Livermore, United States \\ lin32@llnl.gov\
}
\and
\IEEEauthorblockN{Maya Gokhale}
\IEEEauthorblockA{\textit{Lawrence Livermore National Laboratory}\\
 Livermore, United States  \\  gokhale2@llnl.gov\
}
\and
\IEEEauthorblockN{Andreas Herten}
\IEEEauthorblockA{\textit{Jülich Supercomputing Centre}\\
Jülich, Germany \\ a.herten@fz-juelich.de \\
}
\and
\IEEEauthorblockN{Ivy Peng}
\IEEEauthorblockA{\textit{KTH Royal Institute of Technology}\\
Stockholm, Sweden \\ ivybopeng@kth.se\\
}
}
\maketitle

\begin{abstract}
ARM SVE and RISC-V RVV are emerging vector architectures in high-end processors that support vectorization of flexible vector length. In this work, we leverage an important workload for quantum computing, quantum state-vector simulations, to understand whether high-performance portability can be achieved in a vector-length agnostic (VLA) design. We propose a VLA design and optimization techniques critical for achieving high performance, including VLEN-adaptive memory layout adjustment, load buffering, fine-grained loop control, and gate fusion-based arithmetic intensity adaptation. We provide an implementation in Google's Qsim and evaluate five quantum circuits of up to 36 qubits on three ARM processors, including NVIDIA Grace, AWS Graviton3, and Fujitsu A64FX. By defining new metrics and PMU events to quantify vectorization activities, we draw generic insights for future VLA designs. Our single-source implementation of VLA quantum simulations achieves up to $4.5\times$ speedup on A64FX, $2.5\times$ speedup on Grace, and $1.5\times$ speedup on Graviton.
\end{abstract}

\begin{IEEEkeywords}
Quantum simulator, VLA, ARM, SVE, qsim
\end{IEEEkeywords}

\section{Introduction}
Vector-length agnostic (VLA) architectures are emerging as an alternative to vector-length specific (VLS) vectorization on high-end computing platforms equipped with ARMv8-A, ARMv9-A, and RISC-V processors~\cite{stephens2016armv8,gomez2023hpcg,armv9,arm-sve,minervini2021vitruvius}. Today, Intel's SSE and AVX are the most widely used SIMD vectorization for improving computing throughput. Enabling these VLS vectorization requires expanding the instruction sets. Consequently, when the hardware implementation of the vector length changes, new ISA extensions are needed, imposing challenges in code maintenance and portability. For instance, high optimized codes using SIMD intrinsics need to be rewritten when moving to hardware with different vector lengths.

VLA architectures overcome this portability limitation by using a fixed ISA to support different vector lengths. The vector length can vary from 128 to 2048 bits on ARM processors with Scalable Vector Extension (SVE)~\cite{arm-sve}, or even up to 16,384 bits on RISC-V processors with vector extension (RVV)~\cite{waterman2014risc,cavalcante2019ara,minervini2021vitruvius}. Nevertheless, only a single source code needs to be maintained. Several data center grade ARM processors support SVE, including Nvidia Grace, AWS Graviton3, and Fujitsu A64FX. This portability across different architecture generations is promising, but its adoption in production applications still remains an open question, requiring an in-depth understanding of compiler support, system ecosystem, and application co-design.

This paper leverages an important workload for quantum computing, as represented by a production quantum state-vector simulator from Google~\cite{arute2019quantum, quantum_ai_team_and_collaborators_2020_4023103}, to answer these questions. By a study across three ARM platforms, we identify general insights for future VLA designs. As the first step, we assess the current compiler support, including GNU and Clang, for VLA auto-vectorization because it is likely the most used approach for an application to exploit vector units. Our results show that although compilers can auto-vectorize some code regions, a negligible performance gain is achieved. By dissecting the assembly, we attribute this to strided vector memory loads, which result in interleaved memory access that is inefficiently supported in hardware. This leads us to use SVE intrinsics instead to exploit the vector units. 

We propose a generic VLA design for quantum state-vector simulations. However, to achieve high performance, simply achieving high vectorization is insufficient, especially in production applications. Special consideration must be taken to mitigate the unique overhead associated with VLA. For instance, although VLA provides predication to support irregular loops, it requires a tradeoff model to decide whether a scalar or a predicated vectorization is more beneficial. The theoretical upper bound of compute throughput is determined by vector length, but the obtainable performance gain in real-world applications often is affected by the memory subsystem, in terms of data access and bandwidth. Therefore, gate fusion becomes an important system-specific optimization that adapts arithmetic intensity (AI) to meet the balance point on the target platform according to the roofline model~\cite{williams2009roofline}.

We implement the SVE design in the state-of-the-art Qsim simulator and evaluate it on five quantum circuits of up to 36 qubits on different ARM platforms, including Grace, Graviton, and A64FX processors. With a single-source implementation, our VLA quantum simulator achieves up to $4.5\times$ speedup on A64FX, $2.5\times$ speedup on Grace, and $1.5\times$ speedup on Graviton. We leverage a lightweight profiler and PMU event measurement to study vectorization activities. By quantifying metrics including average active vector length, instruction reduction ratio, and memory backend stalls, we draw generic insights for future VLA designs. Our contributions in this work are summarized as follows:
\begin{itemize}[leftmargin=10pt]
    \item  We identify inefficiencies in the current compiler support for VLA auto-vectorization in quantum simulations.
    \item  We propose a VLA design for quantum state-vector simulations and optimization techniques including VLEN-adaptive memory access, buffering, and fine-grained loop control.
    \item We provide an implementation in Google's Qsim, and evaluate five quantum circuits of up to 36 qubits across ARM processors including Grace, Graviton, and A64FX.
    \item Our VLA vectorized simulations achieve up to $4.5\times$ speedup on A64FX, $2.5\times$ on Grace, and $1.5\times$ on Graviton and high scalability up to 288 threads on the JUPITER supercomputer node. 
    \item We define a set of metrics and PMU events to quantify vectorization activities and provide insights for future VLA designs.
\end{itemize}

\section{background and Motivations}
\label{sec:bg}
 

\subsection{Multi-Qubit Quantum Systems}
The mainstream techniques of quantum computer simulations can be categorized into state-vector, tensor-network, and matrix-density methods. State-vector simulation is the most widely used, as exemplified by IBM's Qiskit Aer~\cite{qiskit2024} and Google's Cirq. This work focuses on state-vector simulations for its generality. In an $n$-qubit system, a particle's physical state, denoted as $\ket{\psi}$, can be represented by a normalized linear combination in the Hilbert space $\mathbb{C}^n$: 
\begin{equation}
\ket{\psi} = c_{1} \ket{0...00} + c_{2} \ket{0...01} + \cdot \cdot \cdot + c_{2^{n}} \ket{1...11} 
\end{equation}
where $\sum_{i=1}^{2^n}|c_{i}|^2 = 1$. The complex amplitudes $c_{1}, c_{2},\cdot \cdot \cdot, c_{2^{n}}$ describe the particle's superposition state, while the norm square of $c_i$ represents the probability of the quantum state being observed in the corresponding basic state. A simplified form represents the superposition state in a column vector as:
\begin{equation}
    \ket{\psi} = \begin{bmatrix} c_{1}, c_{2},c_{3}, \cdot \cdot \cdot ,c_{2^{n}} \end{bmatrix} ^ T
\end{equation}
Thus, an $n$-qubit system can be represented mathematically in a vector of $2^n$ complex numbers, called state vector. 
When simulating multi-qubit systems, a state vector simulation models each quantum $k$-qubit gate as a unitary matrix $U_{2^k \times 2^k}$ acting on the state vector via matrix-vector multiplication. For instance, the unitary matrix of Hadamard gate is $H = \frac{1}{\sqrt{2}}\begin{bmatrix}
    1 \quad 1 \\ 1 \; -1
\end{bmatrix}.$
The $k1$-qubit gate and $k2$-qubit gate on different qubits can be regarded as equivalent to the tensor product of their unitary matrices. 
\begin{equation}
    U_{2^{k1+k2} \times 2^{k1+k2} } = U1_{2^{k1} \times 2^{k1}} \otimes U2_{2^{k2} \times 2^{k2}}
    \label{eq:tensor product}
\end{equation}

Using the binary representation of indices, the $k$-qubit gate with $c$-th control qubit acting on the $i$-th qubit can be operated on a pair of amplitudes $c_0$ and $c_1$ as follows:
\begin{equation}
    \begin{bmatrix}
        c'_{*1_{c}...*0_i*...*} \\
        \vdots \\
        c'_{*1_{c}...*1_i*...*}
    \end{bmatrix}_{2^k}
    =
    \begin{bmatrix}
        U_{00} \cdots U_{02^k} \\ \vdots \;\ddots\;\vdots  \\ U_{2^k0}\; \cdots U_{2^k2^k}
    \end{bmatrix}
    \begin{bmatrix}
        c_{*1_{c}...*0_i*...*} \\
        \vdots \\
        c_{*1_{c}...*1_i*...*}
    \end{bmatrix}_{2^k}
    \label{eq:apply-gate}
\end{equation}
where $*$ denotes the arbitrary bits (0 or 1), $1_{c}$ indicates the control qubit, applying the gate only when the selected qubit is in state 1. All other gates without the control qubit can be ignored. In total, $2^{n-k}$ independent computation of such $2^k\times2^k$ matrix and vector multiplication is required in a $n$-qubit system applied $2^k$-qubit gates to get the final result, e.g. $2\times2$ for signle-qubit gate. These small and independent computation kernels show good potential for parallelization. 

\subsection{Vector-length Agnostic Architecture and ARM SVE}
Vector-length specific (VLS) vectorization is the most commonly used single-instruction multiple data (SIMD) technique for boosting compute throughput. For instance, SSE uses 128-bit vectors, AVX2 uses 256-bit vectors, and AVX-512 uses 512-bit vectors. VLS vectorization on modern CPUs is supported via extensions to the instruction sets. Therefore, when the hardware implementation of the vector length changes, new extensions need to be added to the instruction set. Consequently, applications using VLS vectorization need to be rewritten when moving to hardware with different vector lengths, limiting their portability on future hardware.

Vector-length Agnostic(VLA) architectures can overcome the aforementioned limitation. VLA architectures keep a fixed ISA but can support different vector lengths. For instance, ARM Scalable Vector Extension (SVE)~\cite{arm-sve} is an ISA on ARMv8-A and ARMv9-A architectures that can implement different vector lengths ranging from 128-bit to 2048-bit~\cite{arm-sve}. Several recent high-performance ARM processors support SVE, i.e., 128-bit for Nvidia Grace, 256-bit for AWS Graviton3, and 512-bit for Fujitsu A64FX. Similarly, the RISC-V Vector (RVV) extension uses a fixed ISA but supports variable-length vectors, e.g., hardware implementations support vector length up to 4,096-16,384 bits~\cite{cavalcante2019ara,minervini2021vitruvius}.

\begin{figure}[bt]
  \centering
    \includegraphics[width=0.8\linewidth]{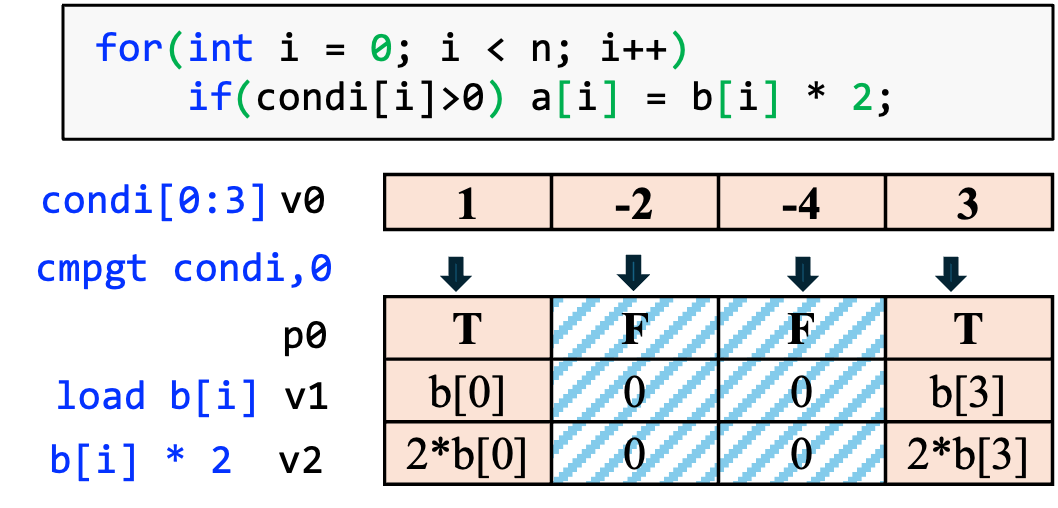}
    \caption{Vectorize a loop with conditional statement using predication to mask out the 2nd and 3rd elements.}
  \label{fig:loop_code}
  \vspace{-0.4cm}
\end{figure}

VLA architectures like ARM SVE and RVV support vectorization of non-uniform strides and conditional statements by using predication to realize fine-grained conditional execution. Loops containing conditional statements, e.g., \texttt{if} and \texttt{switch} as illustrated in Figure~\ref{fig:loop_code}, are challenging for vectorization. The predicate registers in ARM SVE and the masked instructions in RVV can be used to support fine-grained control of each element in a vector register by disabling the corresponding bit in predication or mask. However, identifying profitable loops for vectorization now becomes even more challenging because the overhead of using prediction and the performance gains from vectorized operations need to be weighed. Such reasoning is supported in some compilers' cost models~\cite{pohl2020vectorization}.


\subsection{Qsim Quantum Simulator}


Google's Qsim~\cite{quantum_ai_team_and_collaborators_2020_4023103} is a Schrödinger full state-vector simulator. It is Cirq's backend, used for simulating entropy benchmarking in Quantum Supremacy~\cite{li2019quantum}. Thus, we choose it to represent state-of-the-art quantum state-vector simulators. It is highly tuned to single-precision (FP32) arithmetic, supporting SIMD vectorization on a single core, multithreading on multi-core processors, and acceleration on GPU. The SIMD vectorization is implemented using SSE/AVX intrinsics to support Intel platforms. However, the quantum simulator currently lacks hardware and software co-design to exploit VLA architectures, such as ARM SVE and RVV. This work fills this gap by enabling a high-performance portable quantum simulator across ARM processors.

Qsim provides C and Python interfaces, where the C interface uses manually written input circuits. 
It uses \textit{gate fuser} to analyze the circuit and combines adjacent gates into a single composite unitary matrix, instead of applying each quantum gate individually and sequentially, and is designed to reduce main memory access and improve performance. Currently, greedy gate fusion is implemented in Qsim with a maximum of six fused gates on different qubits~\cite{quantum_ai_team_and_collaborators_2020_4023103}.

\subsection{Motivations} We select Google QSIM for in-depth performance analysis and optimization due to its high impact on the quantum computing community and technical challenges in portable vectorized implementation. The first challenge is effectively vectorizing compute-intensive loops accessing complex number vectors stored in a memory-interleaved format. Second, the requirement for large capacity due to the state space qubit store expansion. Moreover, lessons learned from diagnosing and solving vectorization issues across multiple ARM platforms inform both application developers with similar workloads as well as compiler researchers seeking to extend analysis and transformation to enable automatic vectorization.
\section{Auto-Vectorized Performance on ARM SVE}
\label{sec:auto}
Compiler auto-vectorization is the most widely used approach for leveraging vector units, compared to alternatives, such as intrinsics, macros, or direct assembly programming. It offers a trade-off between programming efforts and vectorization performance. \cite{shi2025arm, lai2025risc} Therefore, we start with evaluating the effectiveness of compiler auto-vectorization for ARM SVE. 

\begin{figure}[bt]
    \centering
    \subfloat[Grace GCC14]{\includegraphics[width=0.33\linewidth]{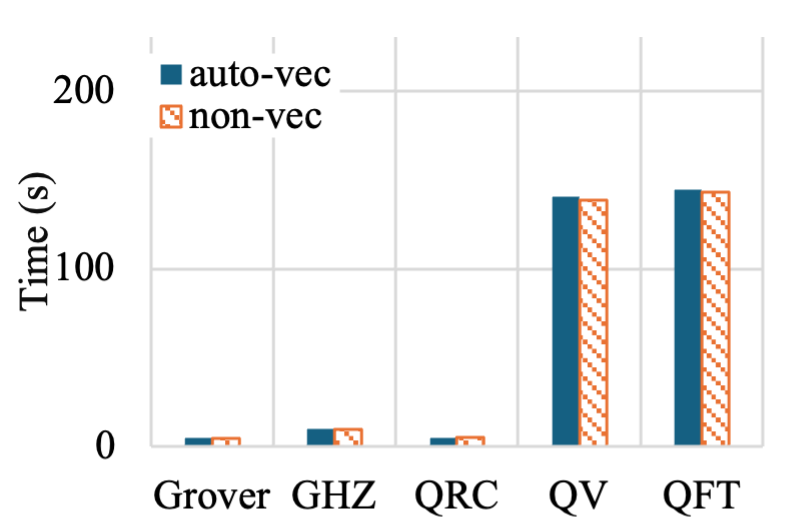}} 
    \subfloat[Grace Clang19]{\includegraphics[width=0.33\linewidth]{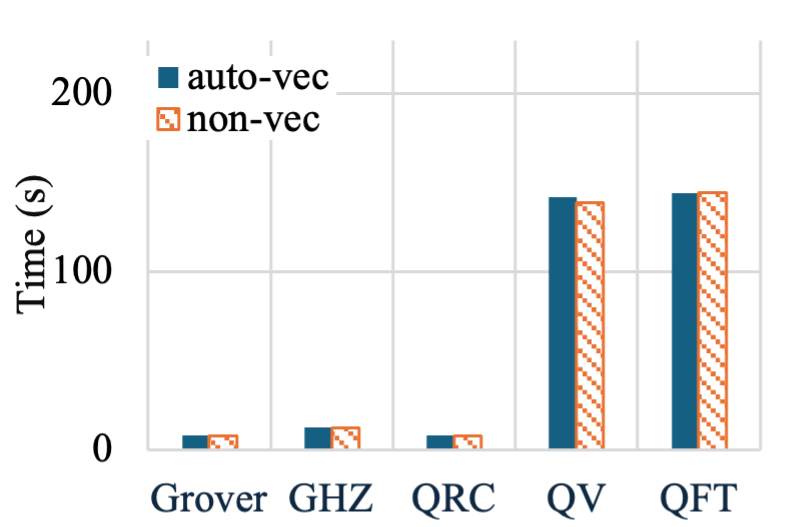}} 
    \subfloat[A64FX GCC12]{\includegraphics[width=0.33\linewidth]{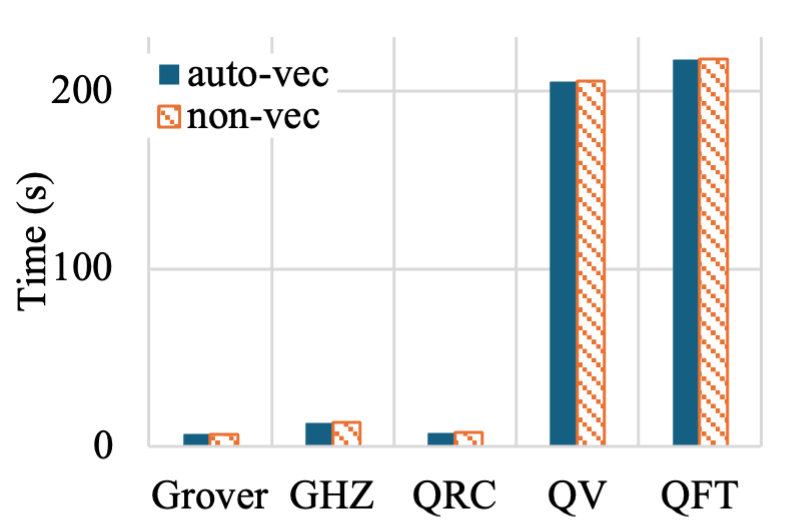}} 
    \caption{The performance of five quantum algorithms on three ARM platforms using VLA via compiler auto-vectorization.} 
    \label{fig:auto_vec}
    \vspace{-0.2cm}
\end{figure}
Figure~\ref{fig:auto_vec} presents the performance comparison of the SVE auto-vectorized version and non-vectorized version on Grace and A64FX platforms. We use architectural flags \texttt{-march=armv8.5-a+sve} in GCC14 on Grace, \texttt{-march=armv8.5-a+sve2 -mllvm -scalable-vectorization=on} in Clang19 on Grace, and \texttt{-march=armv8.2-a+sve} in GCC12 on A64FX to enable the auto-vectorization on SVE. Note that although special compiler flags, such as \texttt{-mautovec-preference} can be used to specify VLA-only vectorization, we choose to use the flags that allow the compiler to make the best choice based on its cost models. Among vector instruction sets, the compiler prioritizes ASIMD over SVE. In GCC’s AArch64 configuration, the number of lanes is compared for both modes, but ASIMD is selected in the case of a tie. Similarly, Clang on AArch64 favors fixed-width vectorization when costs are equal, particularly for epilogue loops or when configured for predictable performance. By examining the compiler report and disassembling, we confirm that auto-vectorization results in the utilization of SVE instructions. Across all five benchmarks on Grace and A64FX by both GCC and Clang compilers, the execution time of the VLA auto-vectorized version is very close to that of the non-vectorized version, i.e., 0.98-1.01$\times$. No obvious performance gain is observed from auto-vectorization. We also measured the VLS vectorized(ARM ASIMD) runtime and no performance gain is shown.

To investigate the behaviors of GCC vectorizer, we disassemble the binary executable and study the assembly in region of interest (ROI). We identify ROI by perf profiling the five benchmarks on Grace. From the functional-level report, we find that the \textit{ApplyGate} function for the common gates takes over 90\% of the execution time, and \textit{ApplyControlGate} and \textit{ExpectationValue} functions dominate the remaining runtime to process special gates like multi-controlled gates and measurements. The critical computation in all these functions is matrix-vector multiplication of complex values to apply gates into the full state vector, as described in Equation~\ref{eq:apply-gate}. When applying the 1-qubit gate, this kernel repeats $2^{n-1}$ times with 28 FLOPs and 24 loads/stores each time. 

\begin{figure}[bt]
    \centering
    \includegraphics[width=0.98\linewidth]{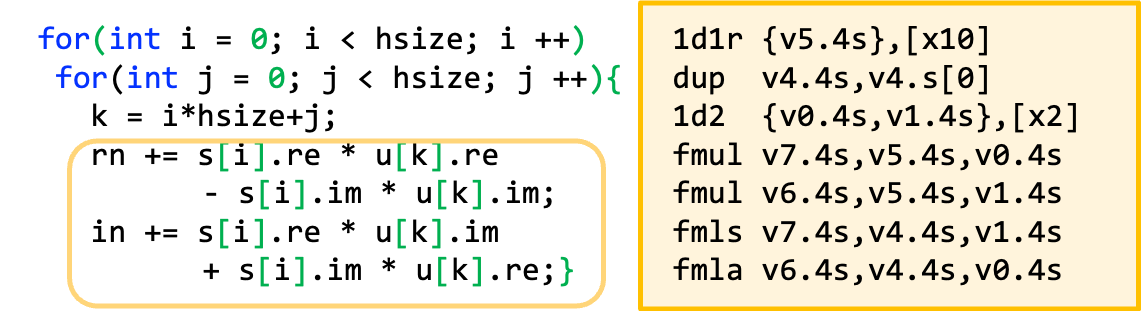}
    \caption{The pseudo code of \textit{ApplyGate} kernel and its disassembly code by compiler auto-vectorization.}
    \label{fig:basic_code}
    \vspace{-0.2cm}
\end{figure}
\begin{figure}[bt]
    \centering
    \includegraphics[width=0.8\linewidth]{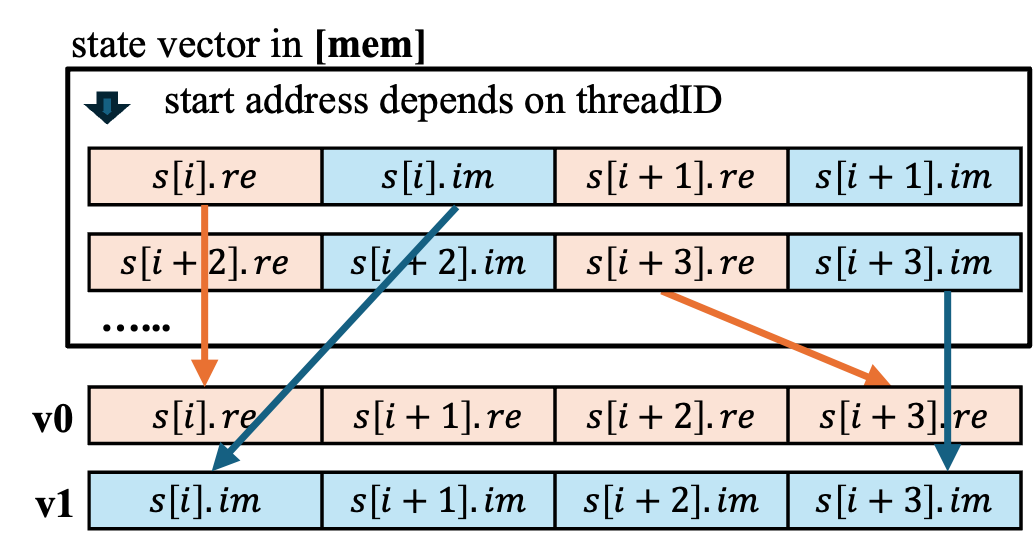}
    \caption{The interleaved memory access pattern in auto-vectorized ApplyGate.}
    \label{fig:basic_mem}
    \vspace{-0.2cm}
\end{figure}

Figure~\ref{fig:basic_code} presents the assembly of the matrix-vector multiplication in an auto-vectorized version. Array \texttt{u} stores the unitary matrices of gates after one-dimensional flattening. \texttt{re} and \texttt{im} represent the real and imaginary parts of the state, respectively. \texttt{ld2} will load with 2-stride and deinterleave pairs of elements from memory into two vector registers~\cite{arm-sve}, illustrated in figure~\ref{fig:basic_mem}. The interleaved memory access pattern results in long-latency load instructions and limits performance gains from vectorization. We further insert \texttt{\#pragma gcc simd} directive in ROI to guide the compiler explicitly, which still results in negligible performance gain compared to the non-vectorized version. Therefore, current compiler support is still immature in supporting VLA architectures and thus, we leverage intrinsics to explicitly utilize ARM SVE in this work.

\section{Vector-length Agnostic Quantum Simulations}
\label{sec:design}
We introduce a general vector-length agnostic design of quantum state-vector simulations. At the high level, state-vector quantum simulators may be parallelized at the thread level in two ways -- parallel computation of state groups or parallel computation of independent gates. If gates are distributed to threads, not all gates can be computed independently. For instance, the controlled gate must wait for the previous operation on the control qubit. To avoid atomic operations and synchronization between threads, partitioning state groups to threads is preferred. 

\begin{figure}
    \centering
    \includegraphics[width=\linewidth]{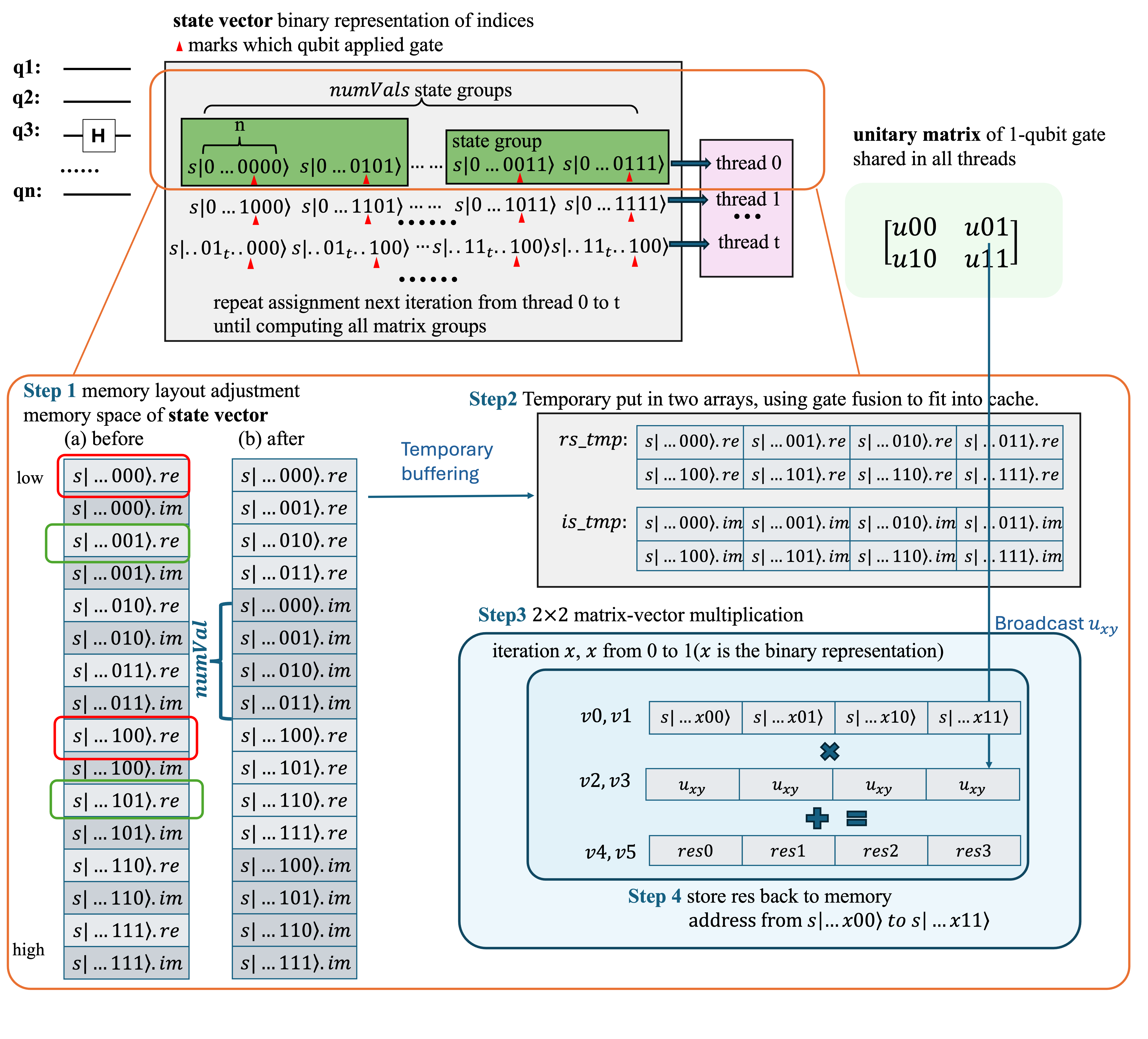}
    \caption{The VLA design of applying gates on an $n$-qubit quantum system. In this example, the Hadamard gate acts on q3, $k=3$, $numVals=4$}
    \label{fig:diagram}
    \vspace{-10pt}
\end{figure}

Parallel computation of state groups (the green box in Figure~\ref{fig:diagram}) distributes the computation of Equation~\ref{eq:apply-gate} to threads, and every gate is computed serially. Unitary matrix of gates is stored in shared memory as a constant array. The indices of the state in a group will keep the same * bits while the $i$-th bit will be 0 and 1 if the gate is applied in $i$-th qubit, as shown in Equation~\ref{eq:apply-gate}. With this layout, the state group parallelism could avoid read and write conflicts. 

Figure~\ref{fig:diagram} illustrates an example of applying the Hadmard gate on the 3rd qubit for an $n$-qubit system.
When assigning the state groups to threads, we iterate the 3rd bit 0/1 and keep the other bits the same within a group. The number of states in a parallel group depends on the gate size, e.g., a single-qubit gate would have two states. Extending the case of a single-qubit gate to a k-qubit gate, the number of states per group is $2^k$. Considering the vectorization, in each thread, we can calculate $numVals$ state groups simultaneously using VLA vectorization. Here, $numVals$ is calculated as $\frac{Vector\; Length (VLEN)}{Element\; Width (ELEN)}$, where $VLEN$ is an architecture characteristic and $ELEN$ is usually 32-bit for single-precision and 64-bit for double-precision. $numVals$ represents the maximum number of data elements that can be operated together in one vector register.


\textit{ApplyControlledGate} and \textit{ExpectationValue} can use the similar loop pattern as \textit{ApplyGate} to compute the matrix-vector multiplication. Special considerations for controlled gates come from the fact that only the indices of states with 1' in the controlled bit will be processed, and others will be disregarded. Therefore, this conditional computation can be supported by using the predication to only keep the state value if the corresponding bit in the current state group's indices is set to 1 (see the conditional statement before step 3 in Figure~\ref{fig:diagram}). To get the final expectation value of selected states, we choose to sum up the magnitude of each state ($\frac{1}{2^n}\sum_{2^n}^{i=0}{||s_i||}$), instead of storing final states back to memory (step 4 in Figure~\ref{fig:diagram}) to avoid long store operations.

To achieve high end-to-end performance, we propose optimization techniques including vectorized loops, memory layout transformation, and gate fusion as follows. 

\subsection{VLEN-adaptive Memory Access Optimization}
Due to the complex datatype of each state, as shown in Figure~\ref{fig:basic_mem}, the memory access for the state vector is interleaved for matrix-vector multiplication. Our evaluation of auto-vectorization in Section~\ref{sec:auto} highlights the importance of consecutive memory access for performance. Several approaches can be employed to address this. First, SVE and many VLS architectures support a vector load instruction with a uniform stride to access an interleaved data layout, but the hardware implementation is still immature. Second, a permute instruction~\cite{fatima2021faster, li2019quantum} can be used to shuffle the right order of amplitudes after consecutive loads to the vector register. However, it is not general to VLA, as different vector lengths require different shuffling order. Another potential limitation is the computation units need to wait for the completion of shuffling after loading into the vector register, under-utilizing the instruction pipeline. Another choice is the split format for complex number, separately storing real and imaginary part into two arrays. It is not chosen for two considerations. naive Qsim uses an interleaved format, leading to essentially require rewriting the entire application to switch the format. Second, copying the data arrays to split format would double the memory requirements, which is highly undesirable since quantum simulations are essentially limited by memory capacity.

In this work, we choose to adjust the memory layout in place during initialization and maintain it throughout the gate applying stage. We adjust to continuously store $numVals$ real floats followed by $numVals$ imaginary floats in memory to ensure that consecutive addresses are loaded from memory into the vector register. Compared to a VLS implementation, for a VLA design, we need to ensure $numVals$ is adaptive to different vector lengths on different architectures. Step 1 of Figure~\ref{fig:diagram} illustrates the memory layout before and after adjustment of the first 16 states when $numVals$ is equal to 4. Assuming a gate is applied to the $i$-th qubit $q_{i}$, where $i$ is greater than $log_2(numVals)$, it will be consecutively loaded into the vector registers for real and imaginary parts, and then $numVals$ state groups are computed simultaneously. This VLEN-adaptive memory layout provides more efficient memory access for VLA architectures with requiring only two additional loops out of the size $2^{n-1}$ in the initialization stage for memory adjustment.

\subsection{Buffering Optimization}
\label{sec:buffer}

When applying gates to state groups on each thread, we have two options to update the state vector after computation. The most straightforward way is to load each value of the state needed in the current state groups (the orange box in Figure~\ref{fig:diagram}), use a temporary array to store the result of matrix-vector multiplication and then update the state vector when the whole result has been computed. However, its performance suffers from the long latency for storing the $2^k$ size of results back to the main memory in each thread, creating high aggregated write back traffic. To reduce store stalls, we assign a load buffer to store all states in this group at the initialization stage (Step 2). Then, the state vector can be updated immediately after one value of the results is obtained without the read-after-write conflicts. This method can reduce the delay in state-vector updates through a good store pipeline and improve IPC.

\subsection{Fine-grained Loop Control}
Step 3 of Figure~\ref{fig:diagram} depicts the vectorized loops for multiplication between the unitary matrix $u$ and the state-vector group, which is the core loop in quantum circuit simulation. The loop size is $2^k \times 2^k$ for the unitary matrix $u_{2^k \times 2^k}$ of $k$-qubit gate. For the outer $x$-th iterations, we compute the results for row $x$ of the state-vector group. For the inner $y$-th iterations, the four consecutive values from \texttt{re\_tmp} and \texttt{im\_tmp} buffers are loaded into vector registers, and the $u_{xy}$ is broadcast to all elements of one vector register. Note that all values are complex numbers, and the real and imaginary parts should be calculated respectively. Three vector load/store and two vector arithmetic operations are processed per iteration. With load buffering, this load is mostly read from the L1 cache to the vector register.


One challenge in efficient vectorization is to identify loops that will improve performance by vectorization, but avoid those that will degrade performance. Ideally, with vectorization, we can reduce the size of loops by $numVals$ times. When applying gates on qubit $q_i$, where $2^i$ is larger than $numVals$, no predicate elements are needed, i.e., they are \textit{regular loops} shown in Figure~\ref{fig:diagram}. However, if $2^i$ is smaller than $numVals$, they become \textit{irregular loops}. The two consecutive states are included in the same state group in \textit{irregular loops}, but only the state in different groups can be processed in parallel. Using the predicate in Figure~\ref{fig:loop_code}, we process these states in sequence. We still load $numVals$ consecutive values, but mask out unnecessary states one by one. When vectorizing an \textit{irregular loop}, if there are more than two active elements in the predicate register, vectorization may still gain performance. However, if the predicate enables only a single active element, it is more efficient to use the scalar unit. We analyze all gates and find the only special condition is multi-qubit gates acting on lower qubits ($q_1$ to $q_{log_2(numVals)}$), will process only one active element per vector instruction. 

\subsection{Gate Fusion Optimization}
\label{sec:gate fusion}
In state-vector quantum simulations, gates acting on different qubits can be fused horizontally by calculating the tensor product of their unitary matrices. Or, gates acting on the same qubits can be fused vertically by the inner product. Vertical fusion combines the gates on the same qubit by $u1_{2^{k1}} \times u2_{2^{k1}}$, which will benefit all platforms at the algorithm level by directly reducing the number of gates without any changes in the unitary matrix size. Horizontal fusion combines the multiple independent gates acting on different qubits by Equation~\ref{eq:tensor product}, which will increase the size of the gate unitary matrix exponentially. In our design, the vertical fusion will greedily fuse all compatible gates, and the horizontal fusion heuristic intends to fuse more gates for improved compute intensity while still keeping the formed matrices in cache. In the following, we refer to the gate fusion parameter $f$ in horizontal fusion. The arithmetic intensity (AI) of matrix-vector multiplication loops after $f$-fused gate becomes  $\frac{2\times(2^{2f}*3+2^f(2^f-1))}{numVals*2^{f+3}}$ flops/bytes. Theoretically, AI increases as fused gates increase. For instance, $AI\approx1.93$ if $f=3$, a substantial improvement compared to $AI\approx0.43$ without fusion. We assume that the vector load from buffers is not the main memory access. However, the simulation will become memory-bound if the unitary matrix size after gate fusion exceeds the L1 cache capacity for load buffering. Thus, the exact choice of gate fusion depends on both the quantum circuit and the memory hierarchy of the hardware platform.



\section{Implementation}
We use ARM SVE intrinsics in the Arm C Language Extension (ACLE~\cite{arm_sve_c_lang_ext}) to implement the VLA design. We choose the intrinsic-based approach instead of directly writing assembly code for its portability and easy maintenance. Table~\ref{tab:intrinsic} summarizes the most used intrinsic, including vector memory access, arithmetic operations and predication operations. The implementation uses 32-bit data precision. 

\begin{table}[bt]
 \centering
 \caption{A summary of main SVE intrinsic used for vectorization}
 \label{tab:intrinsic}
 \resizebox{\linewidth}{!}{
 \begin{tabular}{c|c}
     \hline
     \textbf{Intrinsics} & \textbf{Description}  \\ \hline
      svld1/ svst1& load/ store vector values from memory with scalar base address \\ \hline
      svdup & broadcast a scalar value into selected elements of a vector\\ \hline
      svmls/a & \makecell[c]{multiply the second and third vector and \\add/subtract the result from the first vector}\\ \hline
      svptrue & set all-true predicate for particular vector pattern and ELEN.\\ \hline
      svcmpeq &  \makecell[c]{compare two integer vectors and return a predicate bit \\that indicates whether the inputs are equal.} \\ \hline
 \end{tabular}}
\vspace{-1em}
\end{table}

ACLE defines a new category of data types called sizeless types~\cite{arm_sve_c_lang_ext} to support VLA programming. \texttt{svfloat32\_t} is a sizeless vector type that holds 32-bit lanes while the number of lanes is determined at runtime by \texttt{svlen\_f32}, instead of a fixed value at compilation time as in VLS. However, sizeless types have limitations in programming because sizeless types cannot be used in declaring an array. To overcome this, two scalar arrays are used to store the intermediate variables that fit in the sizeless type as shown in Listing~\ref{lst:sizeless}. \texttt{size} are equal to $2^n$. Compared to using \texttt{memcpy} from \texttt{*p0} and reordering in the matrix-vector multiplication computation, this approach prefetches into the ordered array \texttt{rs\_tmp} and \texttt{im\_tmp} to avoid the long latency loading during computation.

\begin{lstlisting}[language=C, caption=Temorary buffer for the sizeless type,label={lst:sizeless}, basicstyle=\scriptsize, frame=tb,numbers=left, numbersep=5pt,tabsize=1,xleftmargin=0.05\linewidth, xrightmargin=0pt,aboveskip=1.5pt,belowskip=1.5pt]
fp_type re_tmp[gsize * numVals]; 
fp_type im_tmp[gsize * numVals]; 
for (int i = 0; i < size; i++) {
    svfloat32_t v0 = svld1_f32(pg, re_addr);       
    svfloat32_t v1 = svld1_f32(pg, im_addr);  
    svst1_f32(pg, re_tmp + i * numVals, v0);                
    svst1_f32(pg, im_tmp + i * numVals, v1);                
}
... // compute matrix-vector multiplication 
\end{lstlisting}



For loops with the non-uniform strides in irregular memory access, as shown in Figure~\ref{fig:basic_code}, explicit porting with predicate settings is needed. Listing~\ref{lst:predicate} illustrates the algorithm to setting predicate register with the time complexity $O(L)$, where \texttt{L} represents the count of gate-applied qubits lower than $numVals$. \texttt{qs} stores indices of acted qubits. The \texttt{idex} generates a sequence of sequences with a step size of 1 starting from 0 and then it operates bitwise AND with the bitmask \texttt{tmp} to generate the value of predicate \texttt{pg}.

\noindent

\begin{minipage}{\linewidth}
\begin{lstlisting}[language=C, caption=Setting predicate register, label={lst:predicate}, frame=tb, xleftmargin=0.05\linewidth, xrightmargin=0pt,aboveskip=1.5pt,belowskip=1.5pt]
 svbool_t pg = svptrue_b32();
 if(L!=0) {
    uint32_t tmp = 0x0;
    for(int x = 0; x < L; ++x)
        tmp= tmp | (1 << qs[x]);
    svuint32_t idex = svindex_u32(0, 1);
    idex = svand_n_u32_z(svptrue_b32(),idex,tmp);
    pg = svcmpeq_u32(svptrue_b32(),idex,svdup_n_u32(0));
 }
\end{lstlisting}
\end{minipage}


\section{Experimental Setup}
\label{sec:setup}

We use four ARM platforms (Table~\ref{tab:Platform}) to assess the portability across NVIDIA Grace, AWS Graviton3, and Fujitsu A64FX processors. The Grace CPU is based on ARM Neoverse-V2 and two testbeds are used: a single Grace Hopper node with 576~GB memory (the default Grace platform) and a node on the JUPITER supercomputer~\cite{jupiter} that consists of four Grace Hopper superchips interconnected by cache-coherent NVLinks to form a four-NUMA system with a total of 864~GB memory. The Graviton CPU is based on ARM Neoverse-V1 and runs in a VM on AWS EC2 instance. A64FX is the world's first processor that implements SVE and was used on Fugaku Supercomputer~\cite{FUJITSU}. The software environment runs Linux kernel 5.14, OpenMP, Python 3.9.18, and pybind11 (v2.13.6) for Python binding to C++. Quantum circuits are constructed and simulated using Cirq 1.3.0.

Five quantum circuits are implemented using the Cirq library~\cite{Cirq_Developers_2025} for evaluation.
These circuits include gates acting on one- and multiple qubits with and without control qubits. They also cover various depths of gates for a comprehensive evaluation.
Quantum Fourier Transform (QFT), a critical part of Shor's algorithm, performs similar transformation as Discrete Fourier Transform to extract the hidden frequency information relative to phase and magnitudes~\cite{coppersmith2002approximate}. It is composed of Hadamard gates for preparing an entangled state, controlled phase rotation, and swap gates for extraction. 
Grover's search algorithm finds an element in an unstructured database using only $O(n^{1/2})$ queries~\cite{grover1996fast}. The searched element is set as the oracle qubit. It uses X gates and Toffoli gates with multi-controlled qubits to construct the oracle and the diffusion operator. The oracle flips the phase of the target state, while the diffusion operator amplifies its probability amplitude.
Greenberger–Horne–Zeilinger (GHZ) circuit generates the maximally entangled quantum state, with one Hardmard gate on the first qubit and followed by a CNOT gate on each qubit~\cite{fritz2012beyond}. The number of gates grows linearly with the number of qubits.
Quantum Random Circuit (QRC) sampling is proposed by Google for the Quantum Supremacy experiment~\cite{arute2019quantum}. It contains parametrized random rotation gates around the X, Y, and Z axes. $depth$ is the number of layers of unitary gates, $depth=64$ in our experiments. 
Quantum Volume (QV) is designed by IBM for measuring the maximum capability of a quantum device to run reliably for square circuits, where $width = depth$~\cite{cross2019validating}. It selects random permutation of qubit pairs with the CNOT gate and random parameterized unitaries on a single qubit. Table~\ref{tab:benchmark} summarizes these circuits and the number of gate operations performed on qubit $q_i$, depending on whether $i \leq numVals$ or $i > numVals$. When $i \leq numVals$, the simulation needs additional iterations to handle irregular memory access with disabled elements (i.e., predication is masked) in every iteration. We verify the correctness of these quantum circuits by comparing the final state vector of the SVE version to the Cirq built-in simulator with the tolerant error $10^{-6}$.

To profile instructions, memory, and stalls in selected kernels, we develop a lightweight profiler based on perf in C/C++. With it, we collect hardware counters provided by the ARM PMU in region of interest. We also define a set of metrics to characterize the utilization of SVE and identify potential efficiency issues. 
Due to the permissions of AWS instance, we are unable to collect hardware counters on Graviton. 

\begin{table}[bt]
\caption{The specification of experimental platforms.\label{tab:Platform}}
    \centering
\resizebox{0.98\linewidth}{!}{
\begin{tabular}{ccccc}
\hline
\multicolumn{2}{c}{\textbf{Specification}}           & \textbf{Grace} & \textbf{Graviton}  & \textbf{A64FX}   \\ \hline
\multicolumn{1}{c}{\multirow{3}{*}{\textbf{Core}}}   & Arch        &Armv9-A  & Armv8.4-A & Armv8.2-A  \\ 
\multicolumn{1}{c}{}                                 & Count       & 72 & 64 & 48\\ 
\multicolumn{1}{c}{}                                 & Vector Unit & 128-bit SVE & 256-bit SVE & 512-bit SVE\\ \hline
\multicolumn{1}{c}{\multirow{4}{*}{\textbf{Memory}}} & L1   & 4.5 MB & 4 MB & 3MB\\ 
\multicolumn{1}{c}{}                                 & L2   & 72 MB & 64 MB & 4$\times$8 MiB\\ 
\multicolumn{1}{c}{}                                 & Shared LLC  & 114 MB & 32 MB & --\\ 
\multicolumn{1}{c}{}                                 & Main memory & \makecell{480 GB LPDDR5}& \makecell{128 GB DDR5} & \makecell{32 GB  HBM2} \\
\multicolumn{1}{c}{}                                 &Memory BW & 380GB/s & 307.2GB/s & 1024GB/s\\ \hline
\multicolumn{1}{c}{\multirow{1}{*}{\textbf{Compiler}}} & GCC version   & 14 & 13 & 12\\ \hline
\end{tabular}}
\end{table}
\begin{table}[bt]
    \caption{The number of gate operations on qubit $q_i$ in a $N$-qubit system. $numVals$ is the number of elements in a vector instruction.\label{tab:benchmark}}
    \centering
    \resizebox{\linewidth}{!}{
    \begin{tabular}{c|c|c}
        \hline 
        \textbf{Benchmark} & \textbf{No. of Gate Ops ($i \leq numVals$)} & \textbf{No. of Gate Ops ($i > numVals$)} \\ \hline \hline
        QFT & $\frac{1}{2}numVals(numVals+3)$ &  $\frac{1}{2}(N-numVals)(N-numVals+3)$ \\ \hline
        Grover & $5*numVals$ & $5*(N-numVals) + 4$ \\ \hline 
        GHZ & $numVals$ & $N-numVals$ \\ \hline
        QRC & $depth*\frac{1}{4}numVals(numVals+11)$ & $depth*\frac{1}{4}N(N - numVals+11)$\\ \hline
        QV & $ \leq \frac{3}{4} numVals(numVals-1) $ &  $ \leq  \frac{3}{4}  N(N-1) $ \\ \hline
    \end{tabular}}
    \vspace{-1em}
\end{table}

\section{Evaluation}
\label{sec:overall}

  

\begin{figure*}
  \centering

  \subfloat[Grover on Grace]{\includegraphics[width=0.19\linewidth]{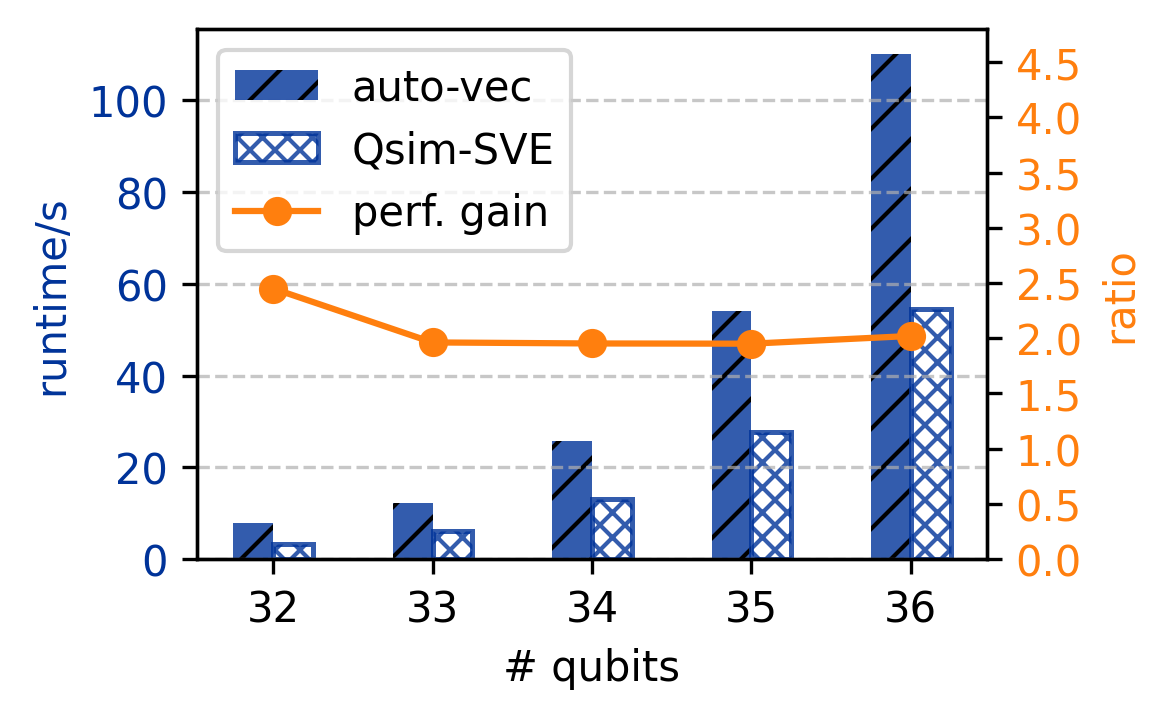}}\;
  \subfloat[GHZ on Grace]{\includegraphics[width=0.19\linewidth]{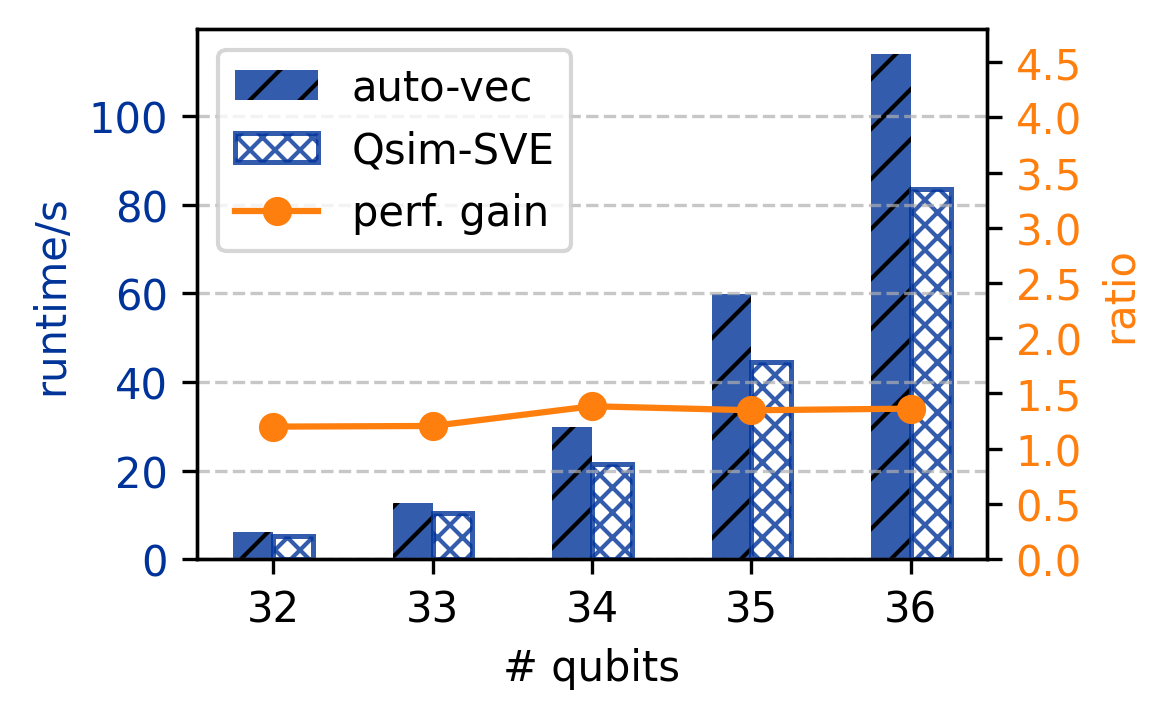}}\;
  \subfloat[QRC on Grace]{\includegraphics[width=0.19\linewidth]{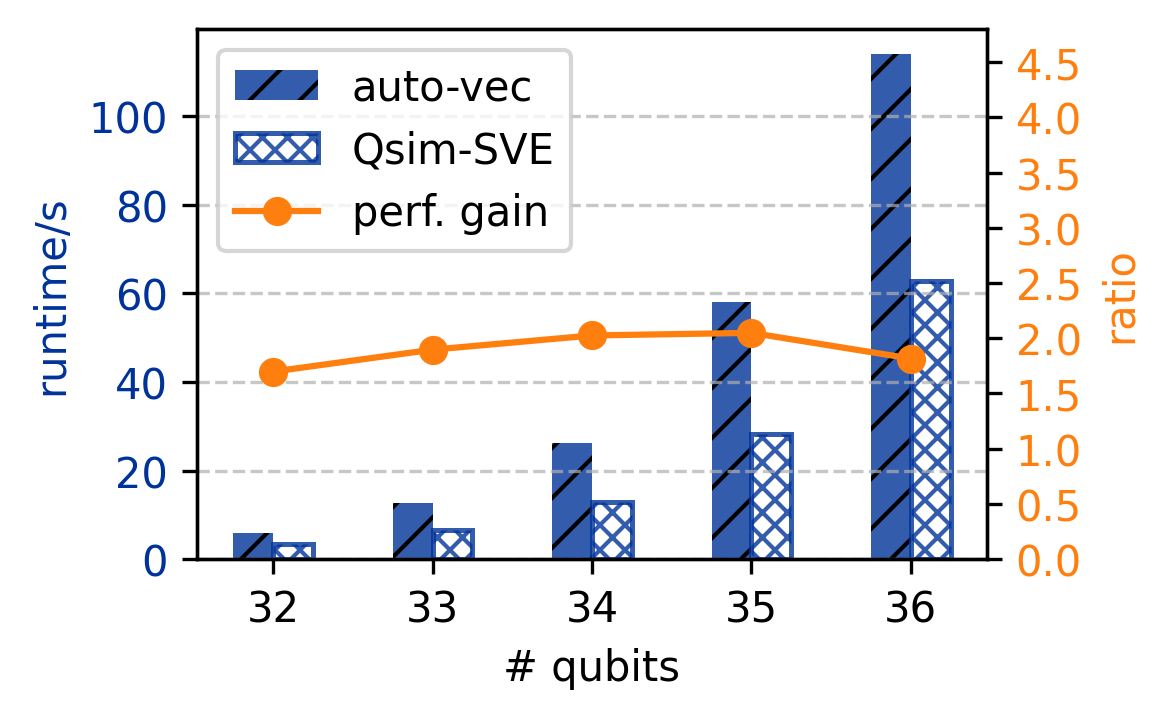}}\;
  \subfloat[QV on Grace]{\includegraphics[width=0.19\linewidth]{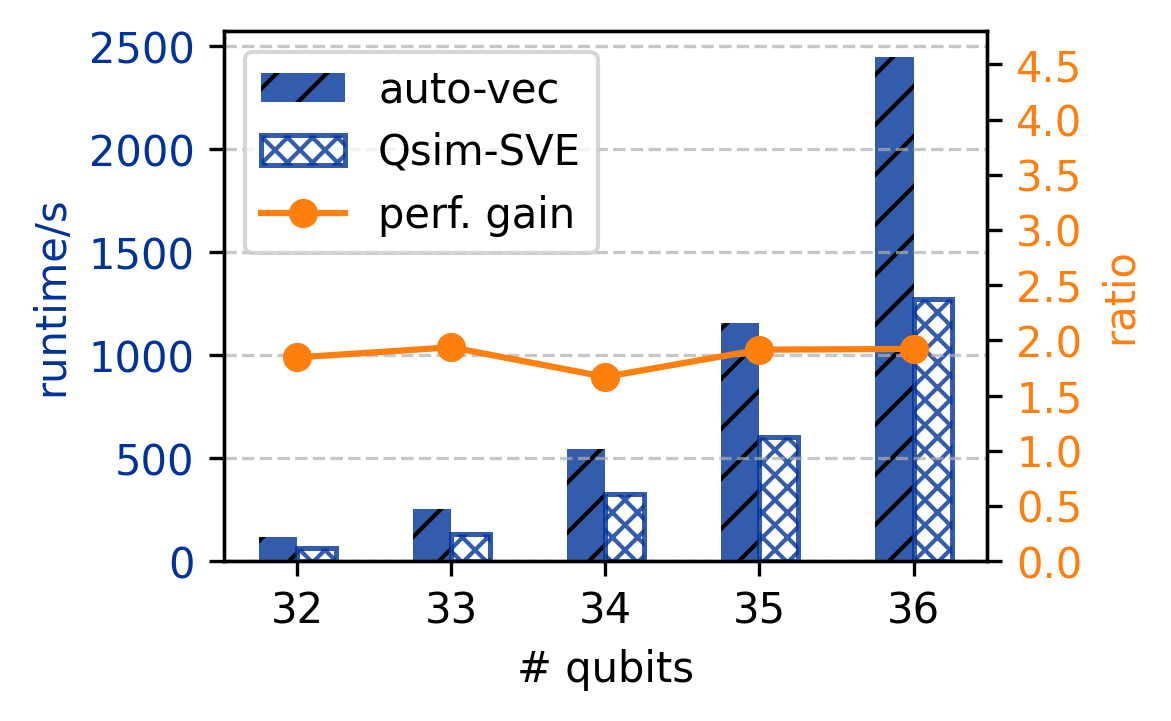}}\;
  \subfloat[QFT on Grace]{\includegraphics[width=0.19\linewidth]{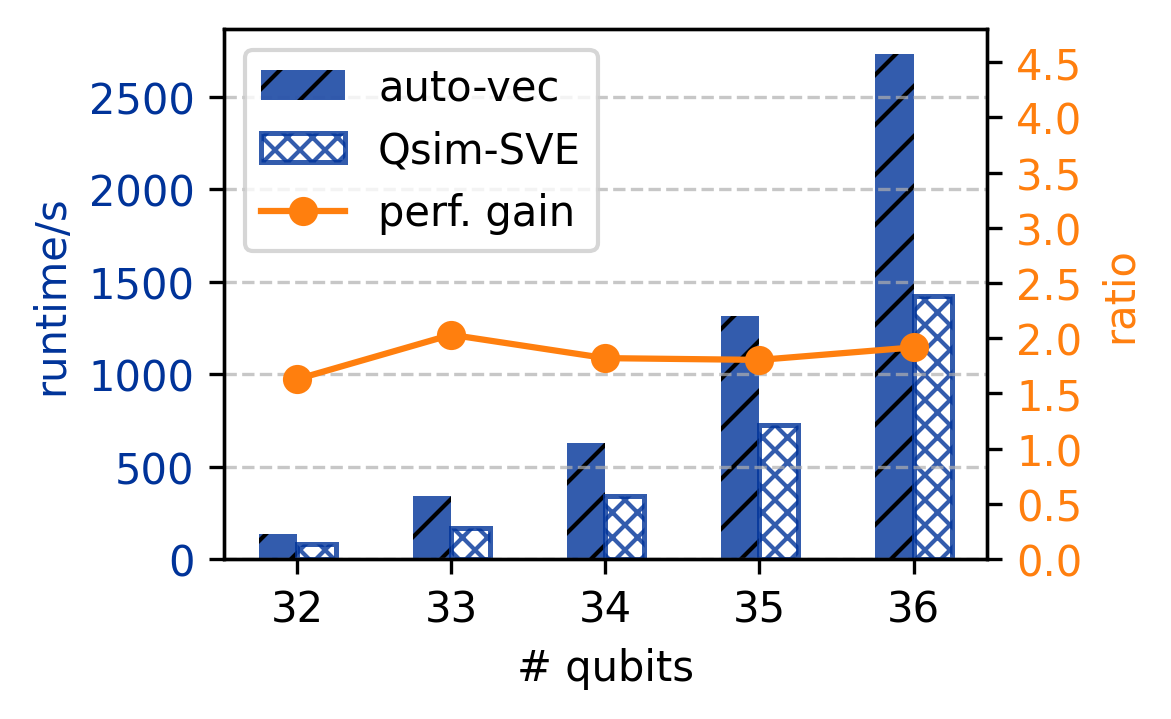}}
\\
  \centering
  \subfloat[Grover on Graviton]{\includegraphics[width=0.19\linewidth]{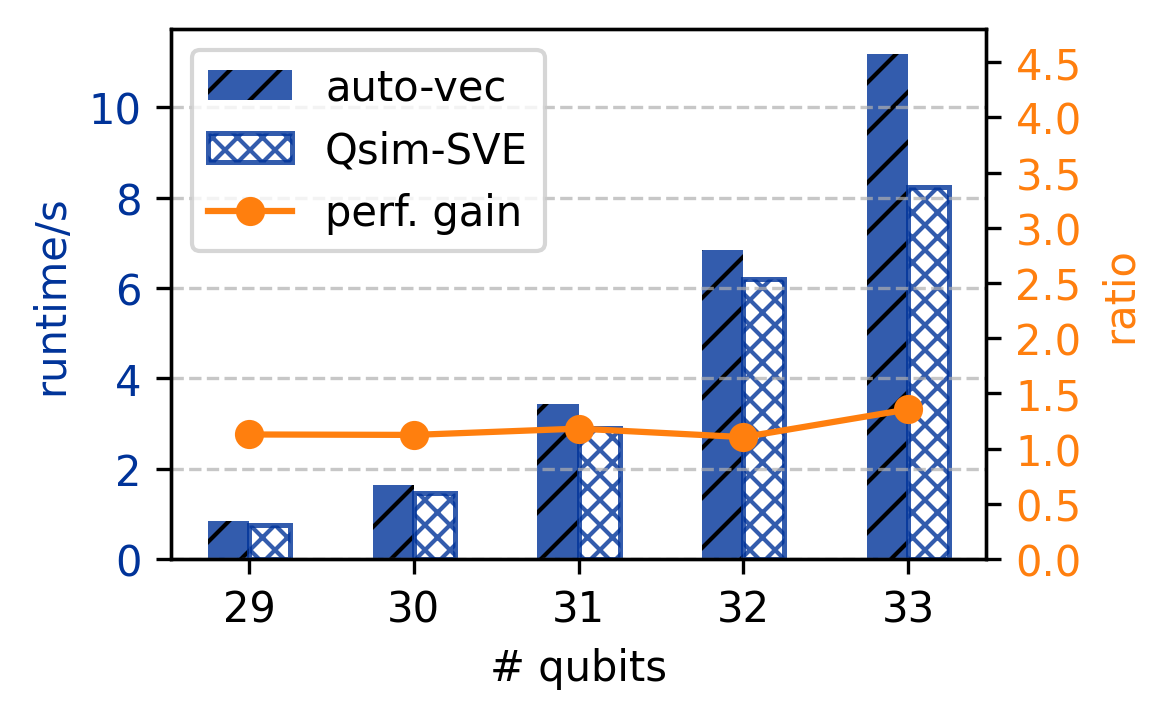}}\;  
  \subfloat[GHZ on Graviton]{\includegraphics[width=0.19\linewidth]{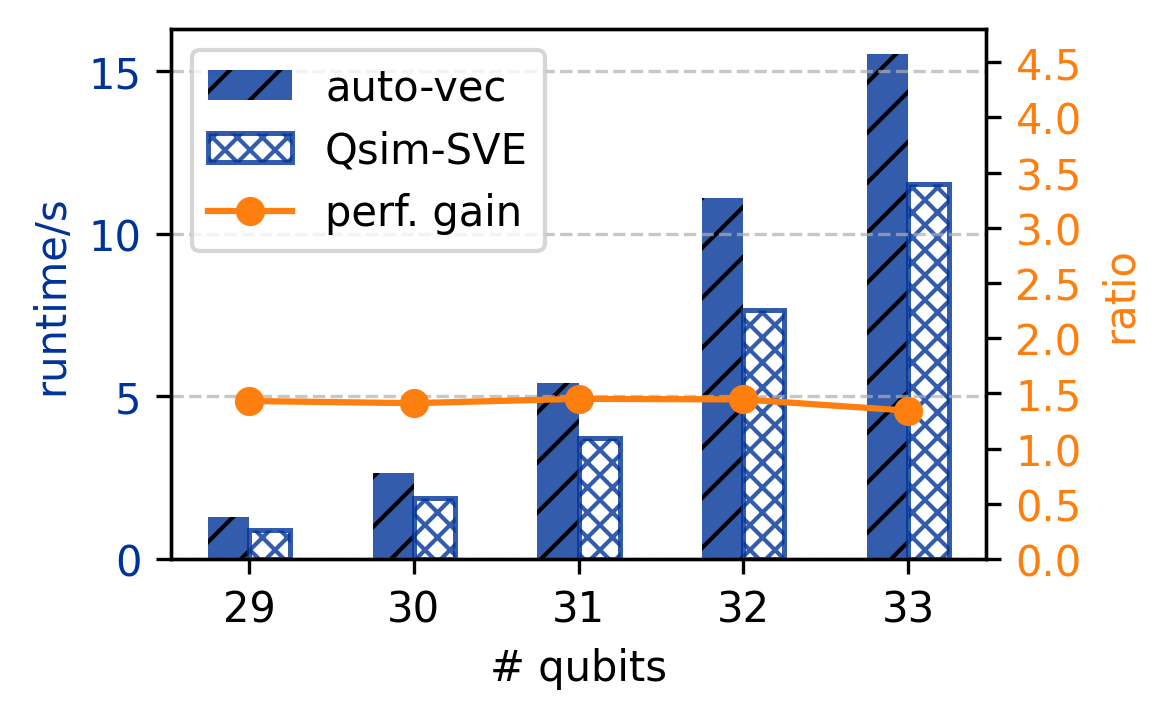}}\;
  \subfloat[QRC on Graviton]{\includegraphics[width=0.19\linewidth]{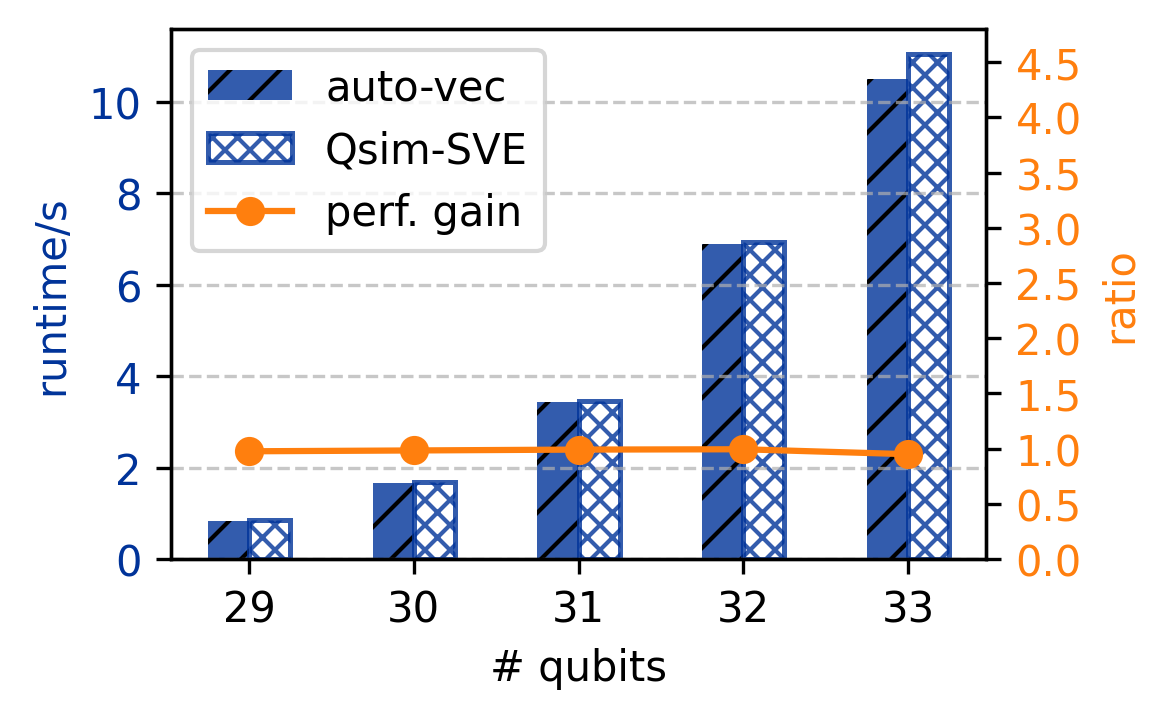}}\;
  \subfloat[QV on Graviton]{\includegraphics[width=0.19\linewidth]{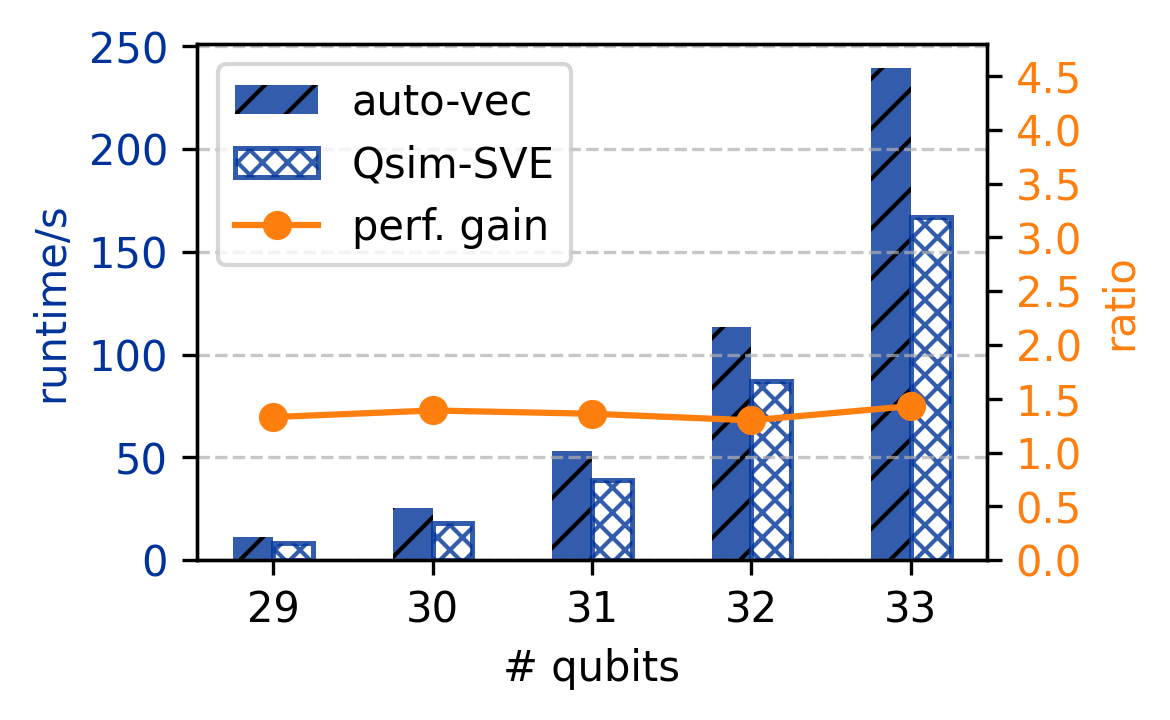}}\;
  \subfloat[QFT on Graviton]{\includegraphics[width=0.19\linewidth]{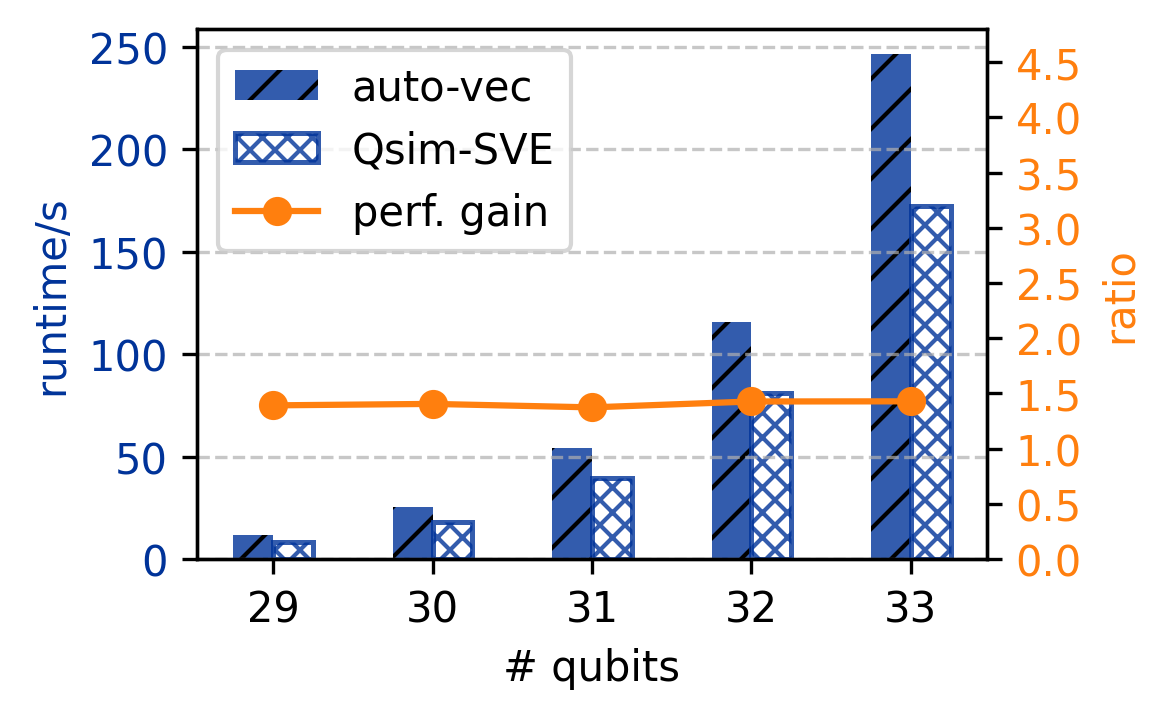}}
 \\

  \subfloat[Grover on A64FX]{\includegraphics[width=0.19\linewidth]{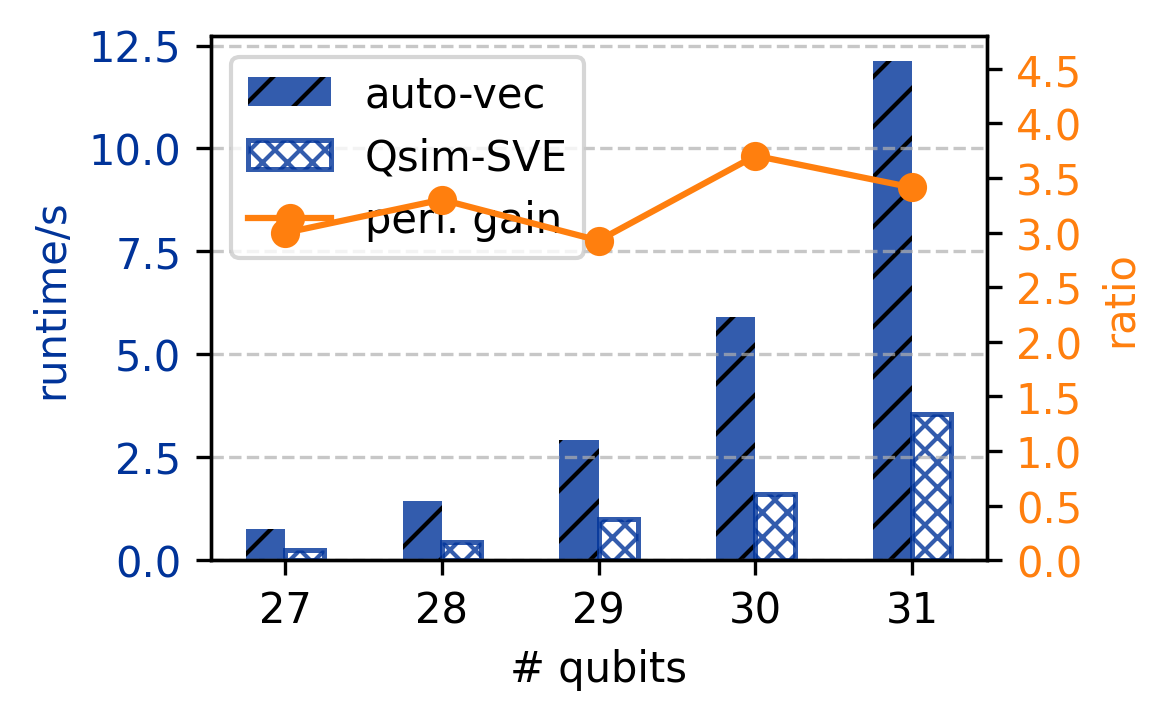}}\;
  \subfloat[GHZ on A64FX]{\includegraphics[width=0.19\linewidth]{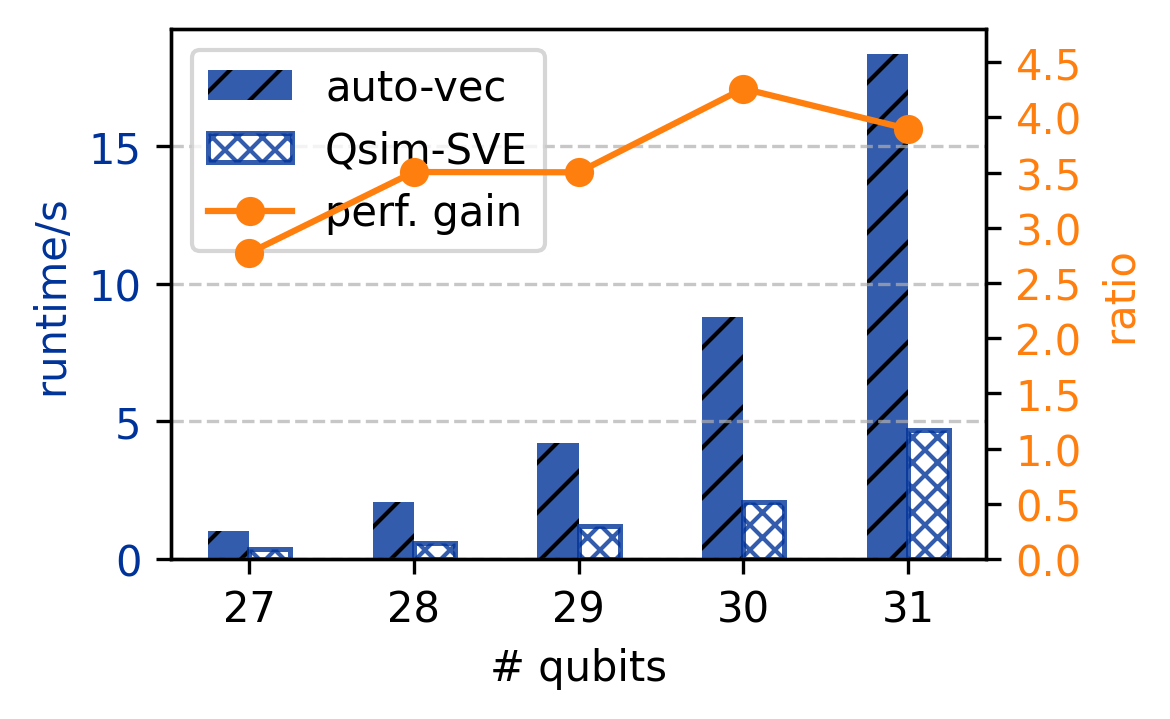}}\;
  \subfloat[QRC on A64FX]{\includegraphics[width=0.19\linewidth]{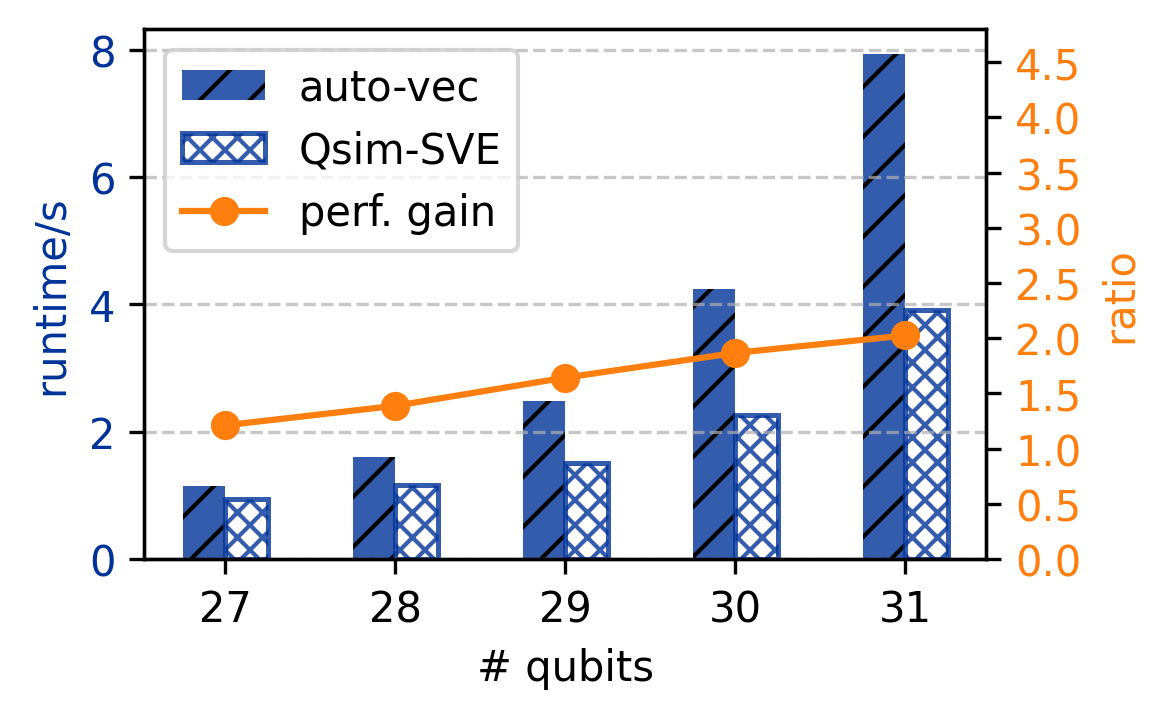}}\;
  \subfloat[QV on A64FX]{\includegraphics[width=0.19\linewidth]{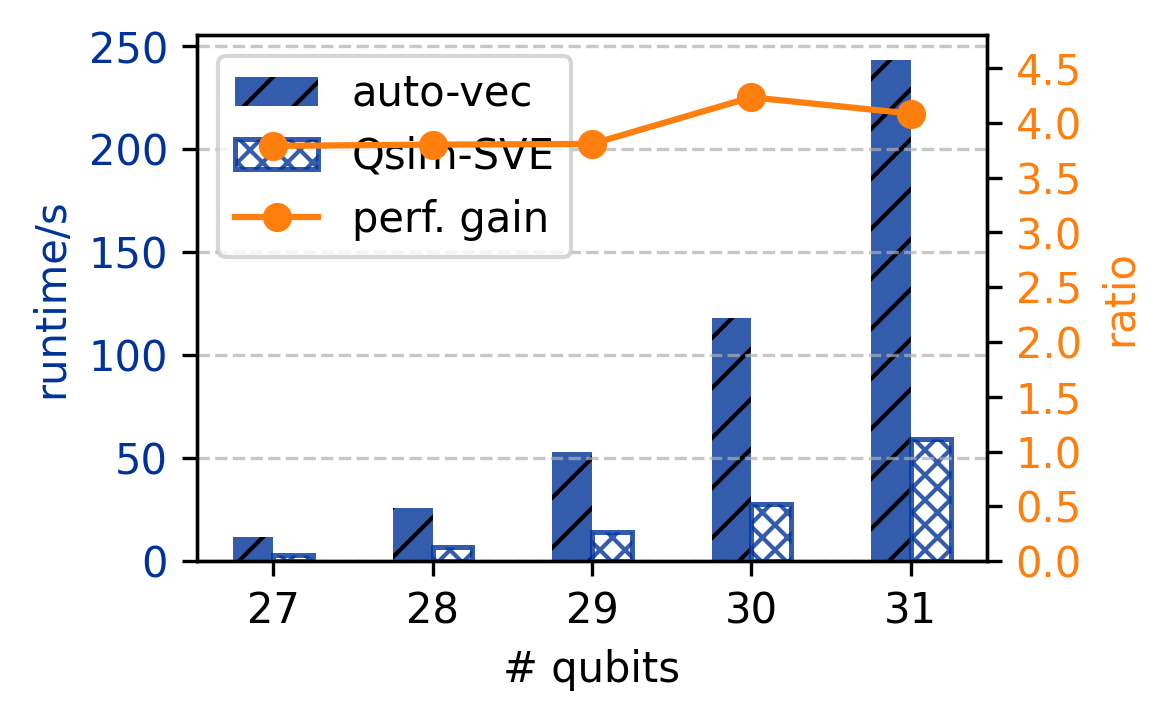}}\;
  \subfloat[QFT on A64FX]{\includegraphics[width=0.19\linewidth]{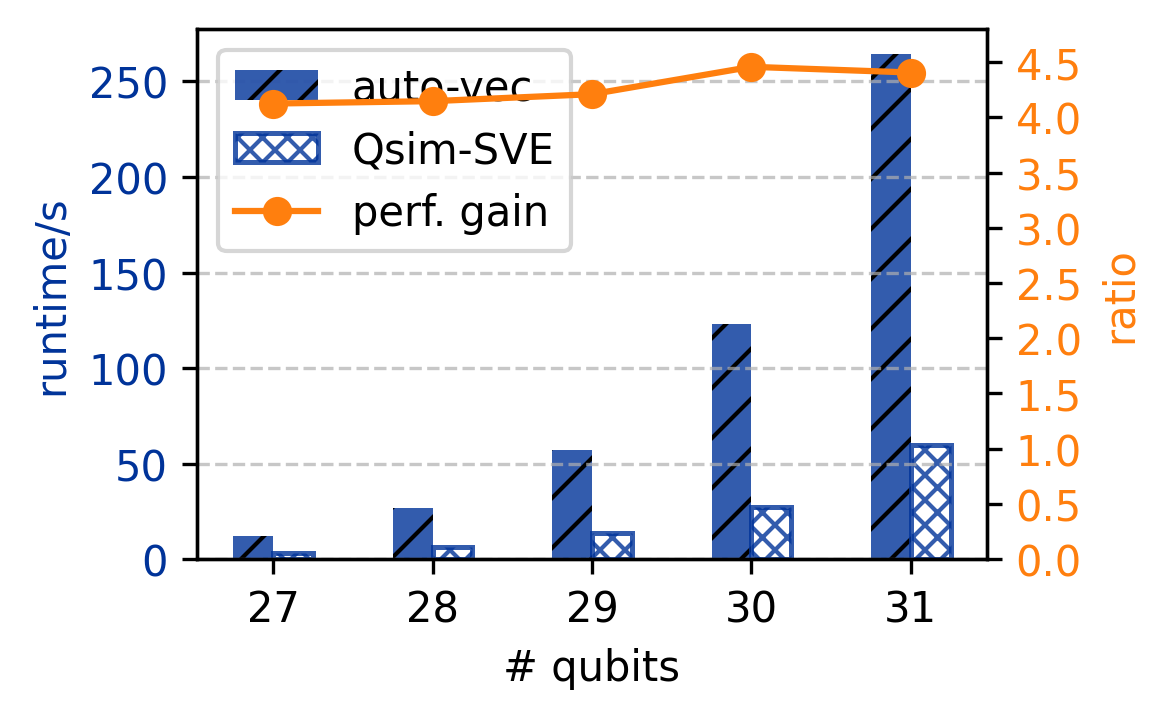}}
   \caption{Overall performance of five quantum circuit simulations using the auto-vectorized version and the SVE optimized version on three platforms equipped with Grace, Graviton, and A64FX, respectively.}
    \label{fig:overall_perf}
\end{figure*}


We start our evaluation with a performance overview of the five quantum circuits on different ARM processors, including Graviton, Grace, and A64FX. For this, we scale up the quantum circuits until the maximum size that can fit in each machine, i.e., 36 qubits on the Grace platform, 33 qubits on the Graviton platform, and 31 qubits on the A64FX platform. On each platform, we use all cores to enable their multi-threaded executions. All circuits on Graviton and A64FX use three fused gates while circuits on Grace use four fused gates except GHZ, which uses three fused gates. To quantify the performance speedup from SVE vectorization, we also run each benchmark using its auto-vectorized version. Fig.~\ref{fig:overall_perf} presents the runtime of the two versions and SVE speedup on all platforms.

The SVE optimized version significantly improves the performance of all quantum simulations on the three ARM platforms, leading to $4.5\times$ speedup on A64FX, $2.5\times$ on Grace, and $1.5\times$ on Graviton. For all quantum circuit simulations, their runtime scales at a doubling rate as the number of qubits increases. This ideal scaling behavior is consistent with $2^{n-1}$ independent matrix-vector multiplication as introduced in Section~\ref{sec:bg}, indicating low overhead in other non-critical computation as the problem size increases. The speedup obtained in the SVE optimization remains mostly consistent for each quantum circuit simulation, even as the number of qubits increases. For instance, the speedup of QFT from 32 to 36 qubits on Grace remains around $2\times$ and the speedup of QV from 27 to 31 qubits on A64FX stays above $4\times$. 


Overall, the SVE optimized version demonstrates high portability across different ARM processors, which cannot be achieved in traditional VLS implementation. Among the three platforms, the SVE speedup on A64FX is the highest compared to Grace and Graviton. The SVE speedup on Grace is consistently higher than that on Graviton, although SVE on Graviton (256 bit) has twice the vector length on Grace (128 bit). This is as expected for two reasons. First, the number of vector units implemented in Graviton ($2\;lanes \times256\;bits$) is the same as in Grace ($4\;lanes \times128\;bits$ SVE), while A64FX implements the longest vector unit of $512\;bits$. Second, shown in Table \ref{tab:Platform}, the memory bandwidth on Graviton is the weakest using DDR4, much lower than that on DDR5 on Grace while A64FX has full high-bandwidth memory (HBM) and provides the highest memory bandwidth. Moreover, accessing Graviton comes with the additional overhead of the virtual machine. Suffering from a limited computational throughput and a suboptimal CPU pipeline, Graviton cannot show its advantage on the longer vector length like A64FX. This finding is important for VLA architectures in general, including ARM SVE and RVV, to consider matching the memory subsystem with the number of functional units, to support high achievable performance at the application level.    


\subsection{Vectorization Activities}
\begin{figure*}
\begin{minipage}[t]{0.3\textwidth}
\includegraphics[width=\linewidth]{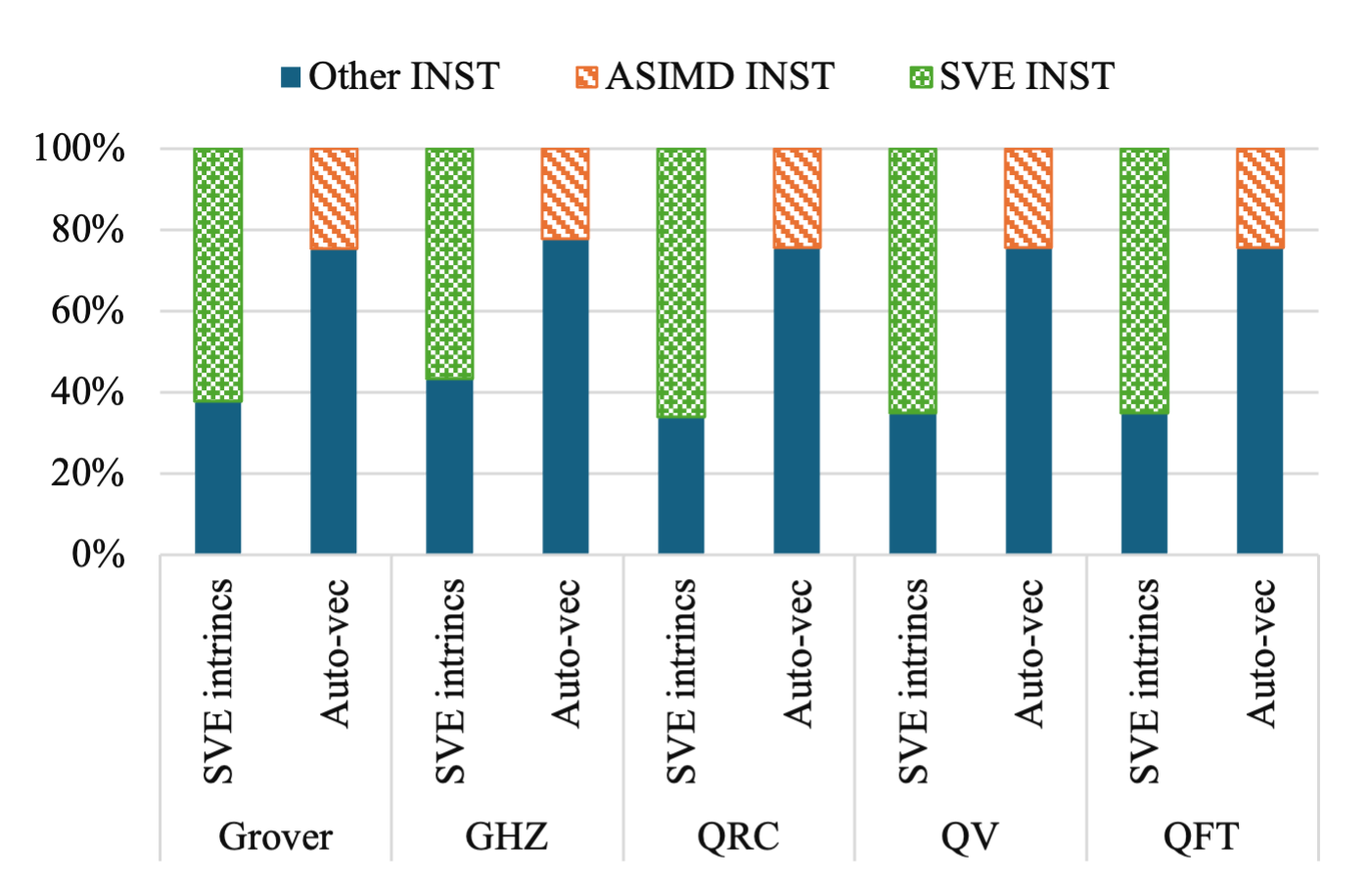}
\caption{Instructions category on Grace.}\label{fig:vector_grace}
\end{minipage}
~
\begin{minipage}[t]{0.33\textwidth}
\includegraphics[width=\linewidth]{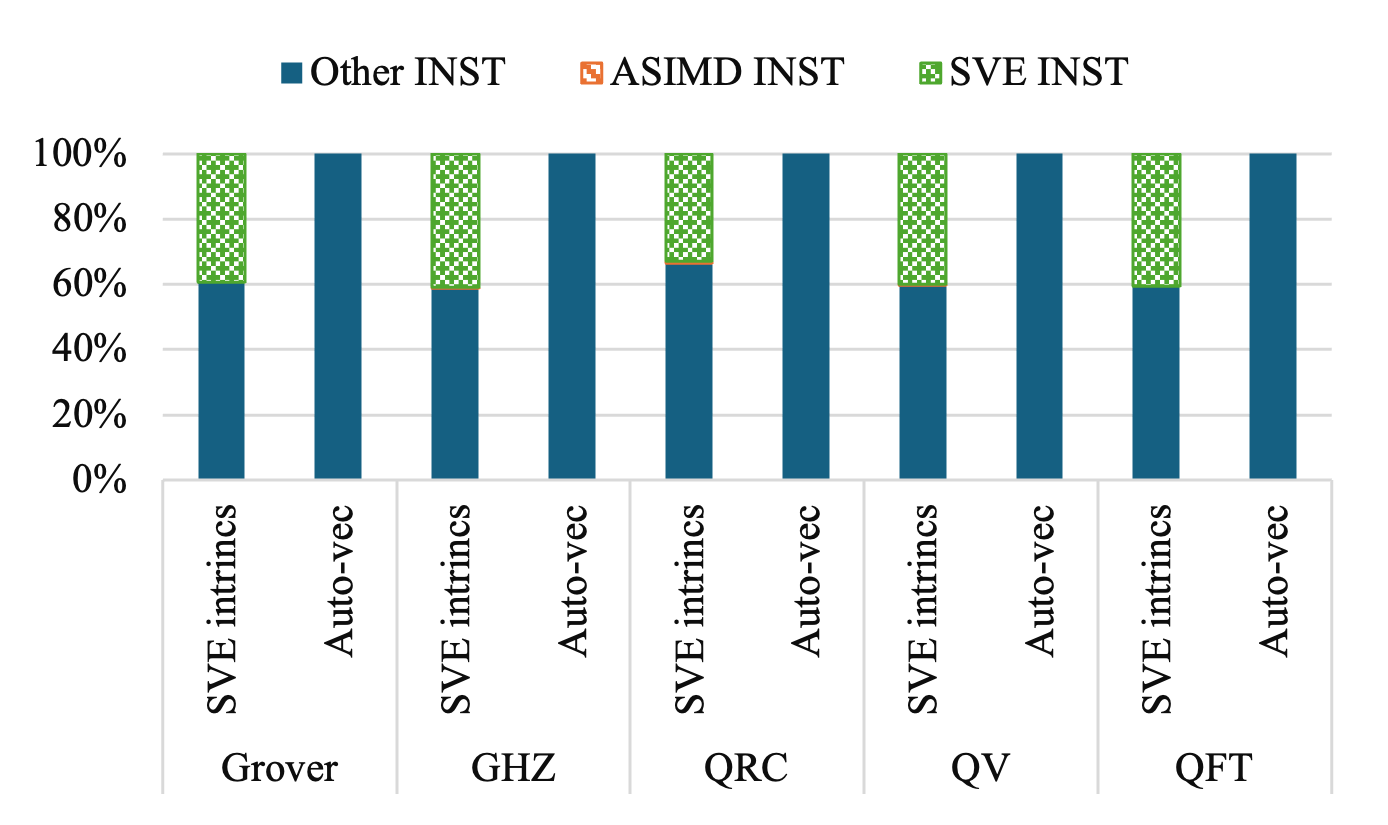}
\caption{Instructions category on A64FX.}\label{fig:vector_a64fx}
\end{minipage}
~
\begin{minipage}[t]{0.36\textwidth}
\includegraphics[width=0.95\linewidth]{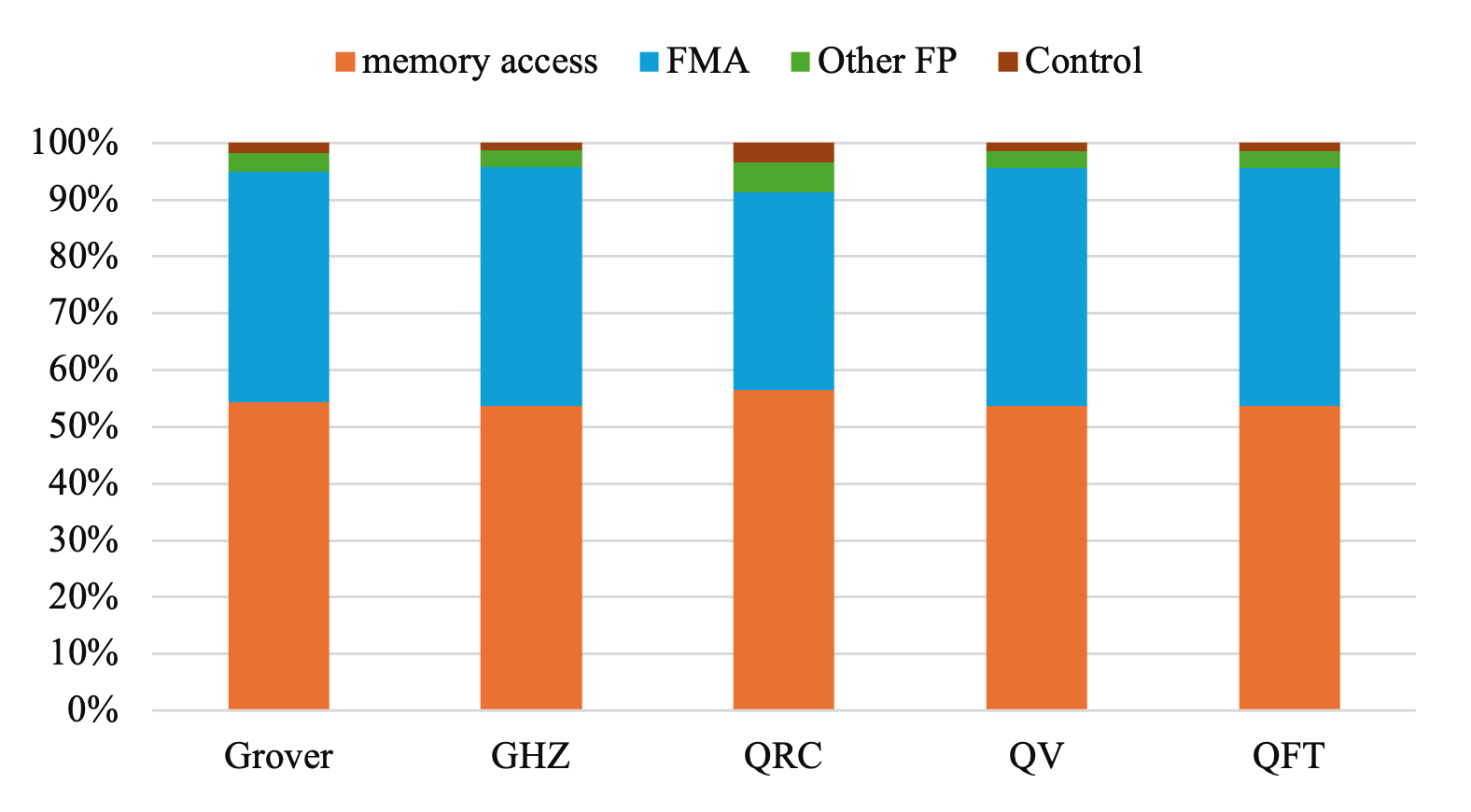}
\caption{Vector instructions on A64FX.}\label{fig:vector_pecentage}
\end{minipage}
\end{figure*}


To understand the effectiveness of vectorization in a production quantum computer simulator like Qsim, we leverage the lightweight libperf library to measure hardware performance counters at the function level. Figures~\ref{fig:vector_grace} and~\ref{fig:vector_a64fx} compare the instruction mixture in the auto-vectorized version and the SVE optimized version. As different ARM processor supports different PMU, we use \texttt{INST\_SPEC}, \texttt{ASE\_INST\_SPEC}, and \texttt{SVE\_INST\_SPEC} on Grace and \texttt{SVE\_INST\_RETIRED}, \texttt{SIMD\_INST\_RETIRED} and \texttt{INST\_RETIRED} on A64FX. By inspecting the ratio of SVE, ASIMD and scalar instructions over the total instructions, the SVE optimization significantly increases the percentage of vector instructions from 20\% to 60\% on Grace, and from nearly 0\% to 40\% on A64FX. The results show that our design can effectively vectorize quantum simulations on ARM SVE units.

We further profile vector instructions to identify SVE instructions critical for quantum simulations. The SVE instructions can be divided into three categories. First, memory accesses use the vector load and store instructions. Second, FP instructions are the SVE arithmetic instructions using FPUs, such as Fused Multiply-Add(FMA). Finally, control instructions is register operation instructions like move, broadcast, and integer operations. Since only A64FX supports relevant PMU events, we profile the following PMU events: \texttt{ASE\_SVE\_LD\_SPEC}, \texttt{ASE\_SVE\_ST\_SPEC}, \texttt{FP\_SPEC} and \texttt{FP\_FMA\_SPEC}. The percentage of memory access, FMA, other FP and control instructions among SVE instructions is presented in Fig.~\ref{fig:vector_pecentage}. Memory access instructions occupy more than 50\% of all SVE instructions, while FP instructions occupy 40\%-45\%. In FP instructions, 86.8\%-93.7\% are FMA instructions. The results highlight the importance of high-performance hardware implementation of memory and FMA instructions in supporting quantum simulations. 

We define a set of metrics to quantify the utilization of the vector units, including average active vector length (AVL) and reduction ratio of retired instructions (IRR), and report them in Table~\ref{tab:avl}.

\textit{Average active vector length (AVL)} measures the number of active elements per vector instructions. It quantifies how effectively SVE instructions utilize VLEN. A similar metric is used in evaluating RVV vectorization~\cite{banchelli2025exploring}. We calculate AVL by weighted average of \texttt{SVE\_PRED\_FULL\_SPEC} and \texttt{SVE\_PRED\_PARTIAL\_SPEC} on Grace. On A64FX, it is computed as the ratio of \texttt{FL*\_VAL\_PRD\_CNT} and \texttt{L1\_PIPE*\_COMP\_PRD\_CNT} over \texttt{FL*\_VAL\_PRD\_CNT} and \texttt{L1\_PIPE*\_COMP}. The SVE optimization achieves high AVL, i.e., 3.5 (out of $numVals=4$) on Grace and 12 (out of $numVals=16$) on A64FX. The results indicate a high utilization of vector registers, although computing lower qubits needs to set the predicate to disable some elements. 

\textit{Instruction reduction ratio (IRR)} measures the reduction of total retired instructions (including SVE and scalar) from the auto-vectorization version to the SVE version. IRR provides a more intuitive view of how SVE can effectively reduce the total number of retired instructions, which is consistent with the SVE speedup in Table~\ref{tab:avl}. For instance, on Grace, the SVE speedup of Grover, QRC, QV and QFT is slightly higher than IRR because extra instructions like bitwise, predicate and integer instructions are introduced when porting to SVE intrinsics. These instructions do not affect the floating-point operations but increase the total number of retired instructions. 

\begin{table}[bt]
    \centering
    \caption{The vectorization activities on Grace and A64FX, using the same configure of overall performance.}
    \resizebox{0.9\linewidth}{!}{
    \begin{tabular}{|c|c|c|c|c|c|c|c|}
        \hline
         &\multicolumn{7}{c|}{\textbf{Grace}} \\ \hline
         \textbf{bench.} & \textbf{IPC} & \textbf{AVL} & \textbf{IRR} & \textbf{speedup} & \textbf{FE} & \textbf{BE} & \textbf{MEM}  \\ \hline 
        Grover & 4.90& 3.75 & 1.5 & 2.45 & 0.3\% & 99.7\% & 40.8\%\\ 
        GHZ & 3.69& 3.67& 1.3  & 1.2 & 0.16\% & 99.9\%& 47.0\% \\ 
        QRC & 5.51&3.43 &1.3  & 1.70 & 0.1\%& 99.9\%& 48.6\% \\ 
        QV & 5.67& 3.40& 1.4& 1.85& 0.15\% & 99.9\%& 39.4\% \\ 
        QFT & 5.40& 3.42& 1.3 &1.63 & 0.13\% & 99.9\%& 40.1\%\\ \hline
         &\multicolumn{7}{c|}{\textbf{A64FX}} \\ \hline
        \textbf{bench.} &  \textbf{IPC} & \textbf{AVL} & \textbf{IRR} & \textbf{speedup} & \textbf{FE} & \textbf{BE} & \textbf{MEM} \\ \hline 
        Grover & 1.36 & 12.3 & 4.5 & 3.4 & 13.7\% & 86.3\% & \textemdash \\
        GHZ & 1.35  & 12.08 & 4.5 & 3.9& 14.2\% & 85.8\% & \textemdash \\
        QRC &  1.55 & 12.3 & 3.5 & 2.0& 4.9\% & 95.1\% & \textemdash \\
        QV &  1.36 & 11.6 & 4.5 & 4.1& 13.7\% & 86.9\% & \textemdash \\ 
        QFT & 1.35 & 12.6 & 4.7 & 4.4& 14.6\% & 85.4\% & \textemdash \\\hline
    \end{tabular}
    }
    \label{tab:avl}
    \vspace{-1em}
\end{table}

While the SVE version achieves high AVL, close to their theoretical limit of $numVals$, the end-to-end performance speedup is much lower. For instance, QFT gets 12.6 AVL but only $4.4\times$ speedup on A64FX. Therefore, we apply the top-down method~\cite{yasin2014top} to identify the performance bottlenecks. Table~\ref{tab:avl} reports the percentage of frontend stall cycles and backend stall cycles over the total stall cycles as FE and BE, respectively. Across five benchmarks, frontend stall is merely 0.09\% - 0.3\% and 4.9\% - 14.6\%, while the backend stall reaches 99.7\%-99.9\% and 85.4\%-95.1\% on Grace and A64FX. Therefore, we further analyzing backend stalls, and find that backend memory stalls occupy about 39.4\%-48.6\%. Since $backend\_bound$ includes $core\_bound$ waiting for drivers and execution ports like ALU/FPU, and also $memory\_bound$ waiting for memory load/store, the results indicate SVE design to be balanced between memory-bound and core-bound.


Finally, we compare IPC on Grace and A64FX. The IPC on Grace ranges from 3.69 to 5.67, whereas the IPC on A64FX ranges from 1.35 to 1.55. Considering the core-bound side, FMA instructions only have two available ports on A64FX~\cite{Fujitsu2020A64FX} with 512-bit VLEN. For applications dominated by FMA vector instructions, as shown in Fig.~\ref{fig:vector_pecentage}, backend stalls increase when FMA instructions are stalled waiting for the available ports, reducing IPC to be less than 2. Therefore, for SVE/RVV designers aiming to improve the efficiency of core-bound workloads similar to quantum simulations, it is desired to have sufficient port support for the vector FP operations when VLEN scales up.

\subsection{Impact of Fused Gates on Arithmetic Intensity}
\begin{figure}[bt]
    \centering
    \subfloat[Grace]{\includegraphics[width=0.33\linewidth]{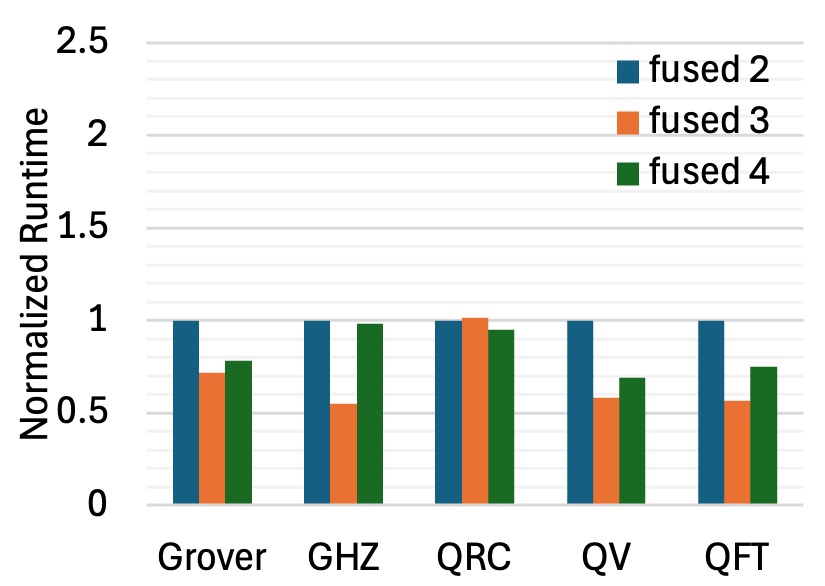}} 
    \subfloat[Gravition]{\includegraphics[width=0.33\linewidth]{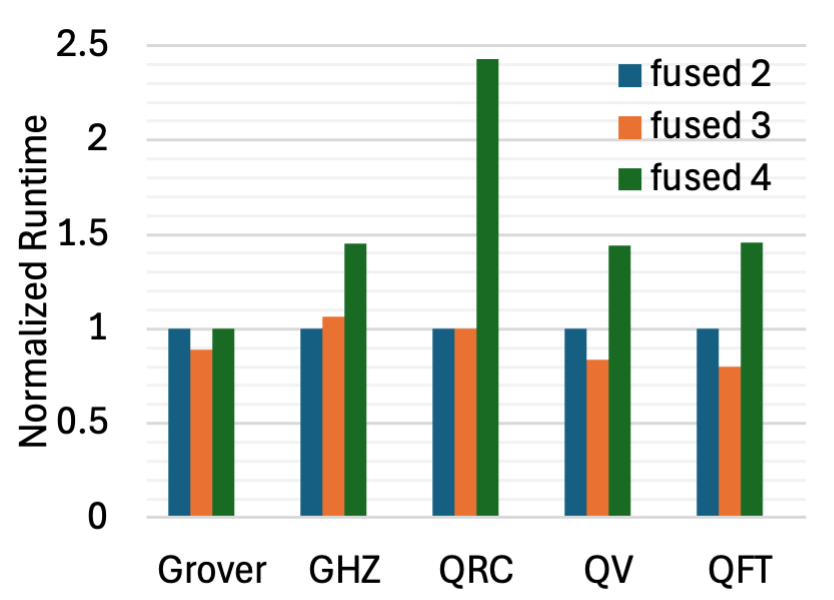}} 
    \subfloat[A64FX]{\includegraphics[width=0.33\linewidth]{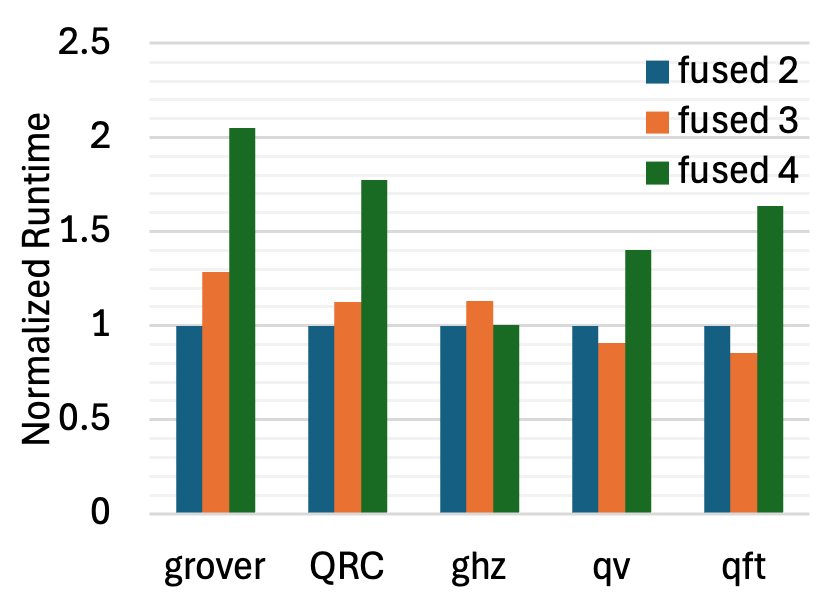}} 
    \caption{The sensitivity test of the maximum fused gates in all benchmarks running on three ARM platforms.}
    \label{fig:sensitivity_fused_gate}
    \vspace{-1em}
\end{figure}
\begin{figure}[bt]
    \centering
    \subfloat[Grace]{\includegraphics[width=0.49\linewidth]{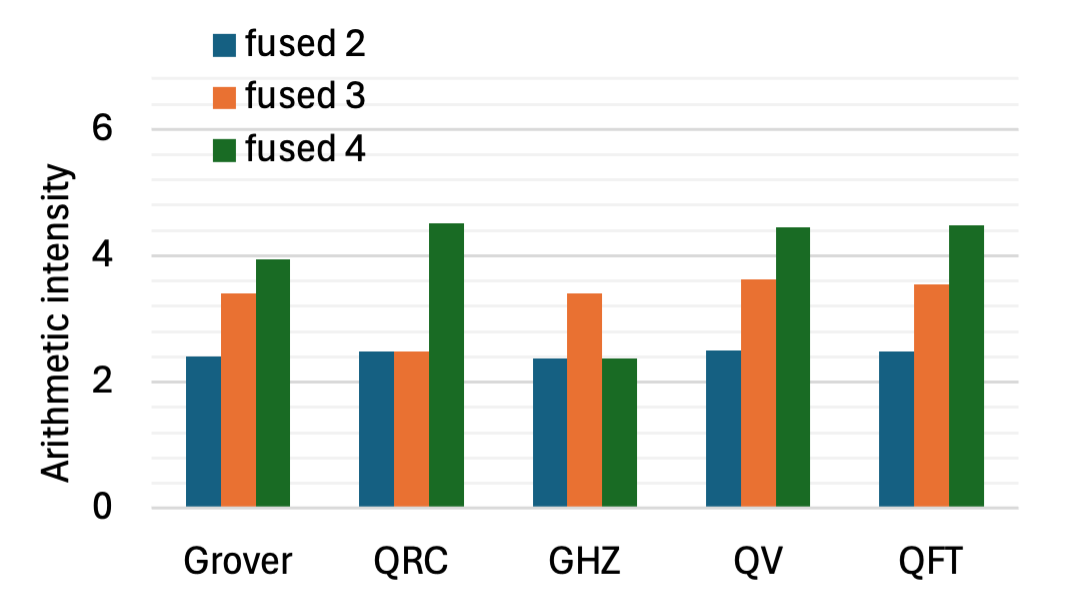}}
    \subfloat[A64FX]{\includegraphics[width=0.49\linewidth]{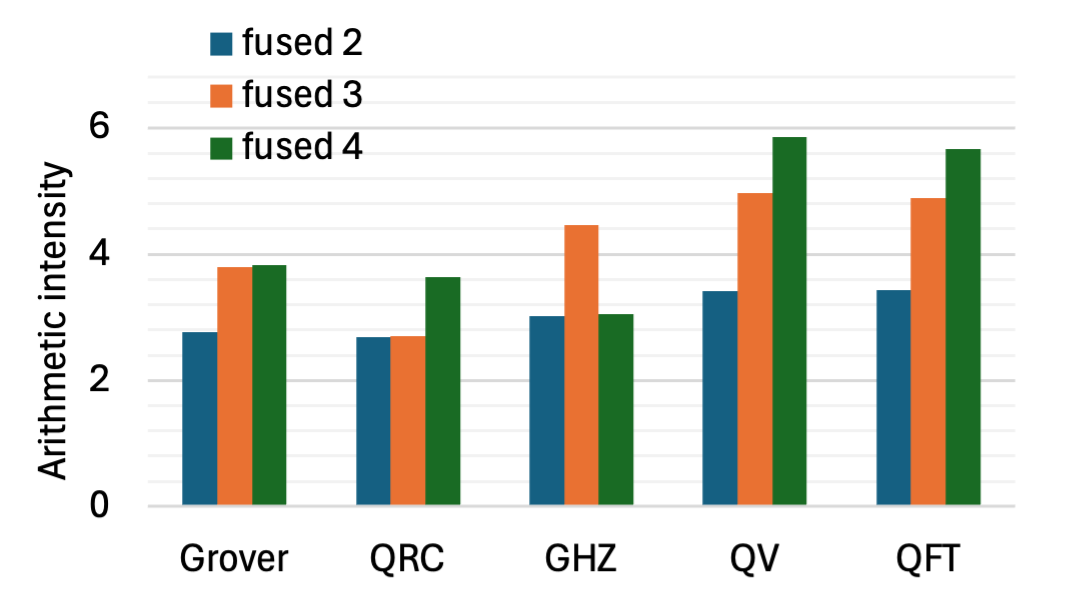}} 
    \caption{The impact of fused gates on Arithmetic Intensity.}
    \label{fig:ai}
    \vspace{-1em}
\end{figure}

Gate fusion is a architecture-specific optimization that adapts Arithmetic Intensity (AI) on a target platform. Fig.~\ref{fig:sensitivity_fused_gate} presents a sensitivity test of fused gates on runtime. For Grover, using two fused gates brings $1.3\times$ speedup over using three fused gates on A64FX. However, three fused gates achieve the best performance on Grace and Graviton. QRC with four fused gates brings $1.8\times$ speedup on Grace, but degrades performance by $60\%$ on Graviton and A64FX. 

We quantify the impact of gate fusion on AI by measuring \texttt{FP\_OP}/\texttt{MEM\_ACCESS}, as reported in Fig.~\ref{fig:ai}. In general, AI increases as the number of gates after gate fusion is reduced, indicating that the workload is gradually shifting from memory-bound to computational-bound. However, the spatial and temporal overhead of temporary load buffering also increases as the number of fused gates grows. Therefore, we propose to choose the gate fusion to make the AI close to the machine balance for good performance. However, the machine balance is determined by many architectural factors, such as the number of cores, CPU frequency, memory bandwidth, and whether SVE is enabled. To ease the search for the optimal fused gate, we propose a synthetic benchmark that has no controlled gates so that gates are reduced linearly with fusion. The synthetic benchmark applies a variety of one-qubit gates on higher qubits (where indices $i> numVal$) to eliminate the impact of circuit structure. The optimal fused gate on Grace found by the synthetic benchmark is three for 32 threads and four for 72 threads, which is consistent with the sensitivity test and overall performance.





\subsection{Breakdown of Optimizations}
\begin{figure}[bt]
    \centering
    \subfloat[Grace]{\includegraphics[width=0.325\linewidth]{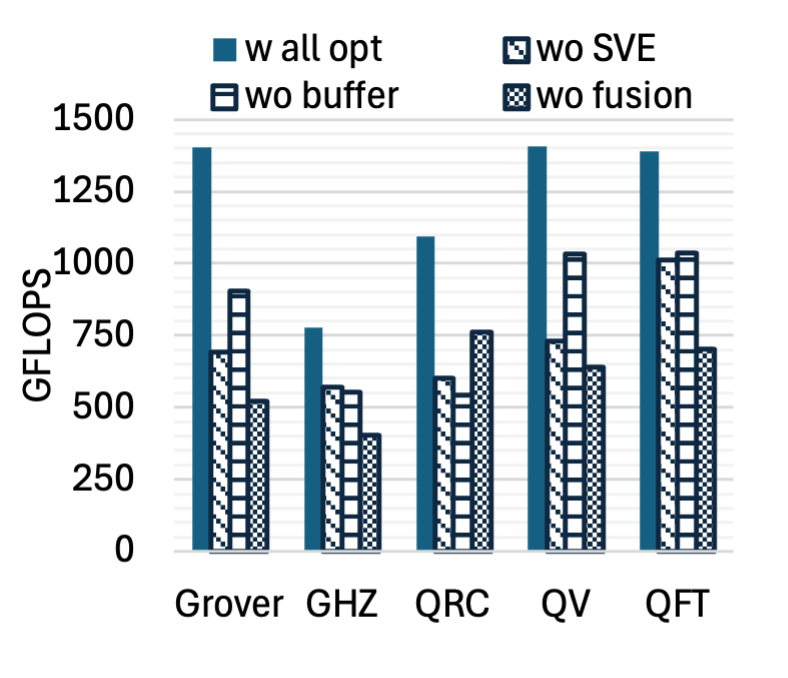}} 
    \subfloat[Graviton]{\includegraphics[width=0.325\linewidth]{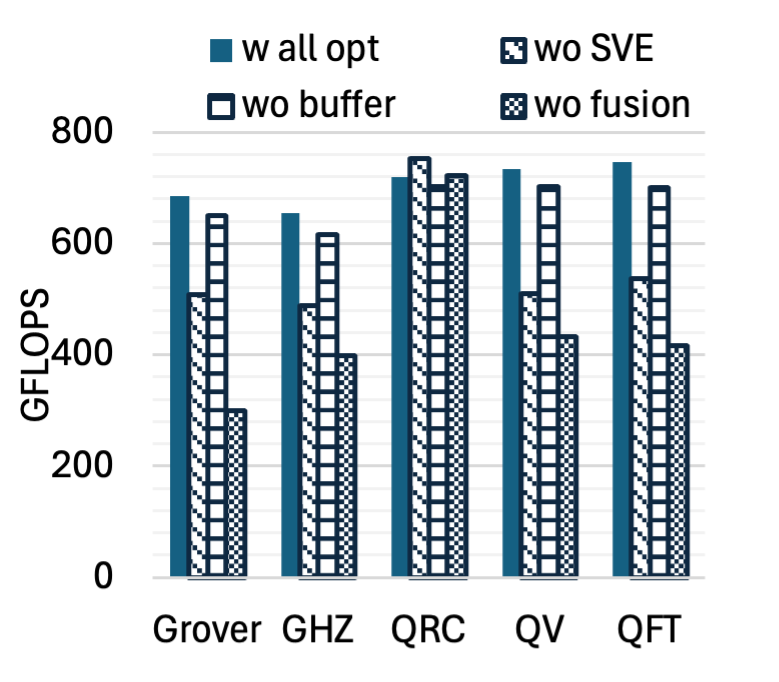}}
    \subfloat[A64FX]{\includegraphics[width=0.325\linewidth]{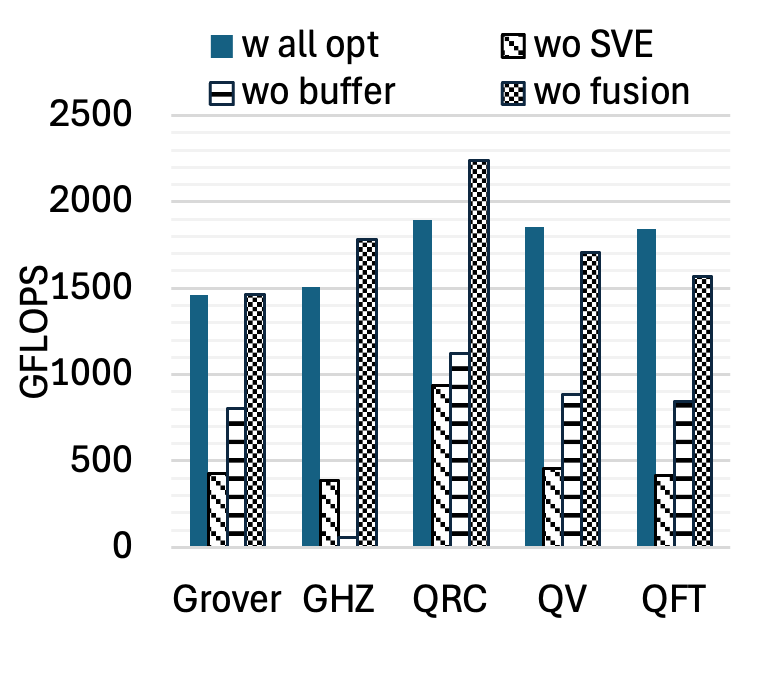}}
    \caption{The ablation test of optimization techniques on all benchmarks on three ARM platforms.}
    \label{fig:opt}
    \vspace{-10pt}
\end{figure}

To evaluate the effectiveness of the key optimization techniques, we perform an ablation test by disabling each individual optimization, including SVE vectorized, temporary load buffer, and gate fusion. From the results  across the three platforms in Fig.~\ref{fig:opt}, SVE vectorization contributes the most performance gain, i.e., 40–50\% performance on Grace and Graviton, or even more profoundly 70-80\% performance on A64FX. Gate fusion also contributes notable performance improvement on Grace and Graviton, improving performance by 20–50\% performance on Grace and Graviton depending on different quantum circuits. However, Gate fusion on A64FX is less effective, where only QV and QFT show a performance gain of 20\%. Finally, the temporary load buffer leads to substantial speedup on Grace and A64FX, but no performance improvement on Graviton, due to the limited 32MiB shared LLC capability on Graviton and lower memory bandwidth compared to the other two platforms. Overall, these optimizations work together to bridge the architectural differences beyond CPU hardware implementation of SVE on different platform, contributing to performance portability.

\subsection{Scalability}
\begin{figure}[bt]
    \centering
    \subfloat[Grace]{\includegraphics[width=0.49\linewidth]{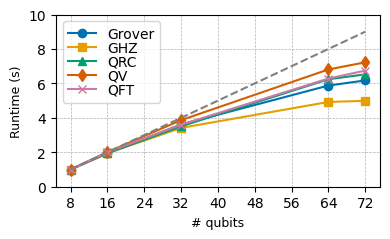}} 
    \subfloat[Graviton]{\includegraphics[width=0.49\linewidth]{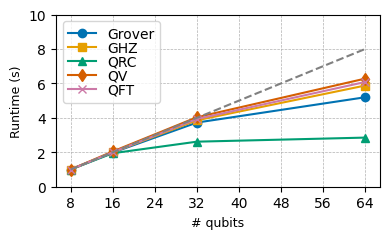}}\\
    \subfloat[Quad-Grace (JUPITER)]{\includegraphics[width=0.49\linewidth]{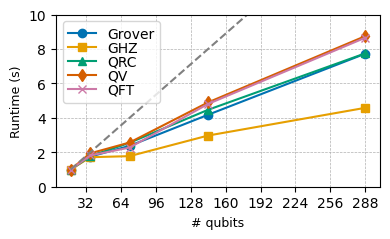}}
    \subfloat[A64FX]{\includegraphics[width=0.49\linewidth]{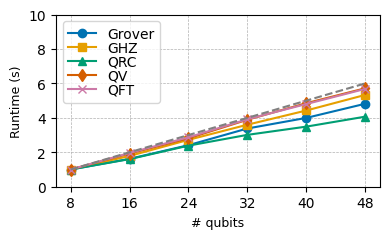}}
    \caption{The speedup of strong scaling with 36-qubit on Grace, 33-qubit on Graviton and 31-qubit on A64FX.}
    \label{fig:scaling}
    \vspace{-10pt}
\end{figure}

\begin{figure*}[bt]
    \centering
    \subfloat[Grover]{\includegraphics[width=0.19\linewidth]{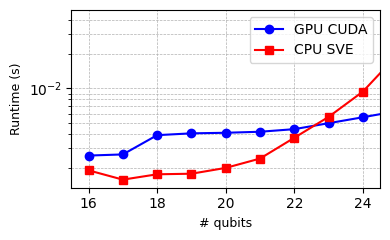}} 
    \subfloat[GHZ]{\includegraphics[width=0.19\linewidth]{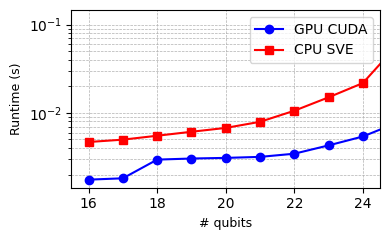}}
    \subfloat[QRC]{\includegraphics[width=0.19\linewidth]{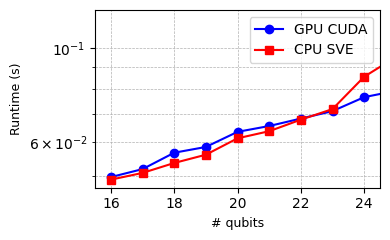}}
    \subfloat[QV]{\includegraphics[width=0.19\linewidth]{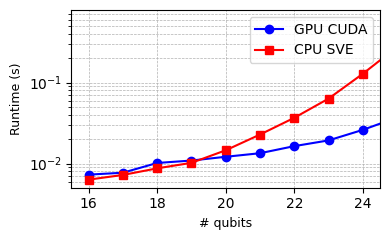}} 
    \subfloat[QFT]{\includegraphics[width=0.19\linewidth]{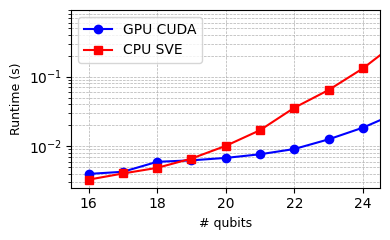}}
    \caption{A runtime comparison of the SVE version with the H100 GPU accelerated version on five quantum circuits.}
    \label{fig:GPU}
    \vspace{-10pt}
\end{figure*}

Fig.~\ref{fig:scaling} presents the strong scaling test using SVE across four platforms, including a four-Grace Hopper node on the JUPITER supercomputer. The benchmarks use the largest problem size on each platform and optimal gate fusion. 
The SVE version exhibits high scalability on five benchmarks as the number of threads increases. For instance, on A64FX, QSim achieves 4.1-5.7~$\times$ speedup on 48 threads relative to 8 threads. Similarly, on Grace and Graviton, QSim achieves up to 7.2~$\times$ and 6.2~$\times$ speedup, respectively, using 72 and 64 threads relative to 8 threads. We note that on A64FX, the scaling performance sustains close to the ideal linear scaling (the dotted line). However, on Grace and Graviton, their scaling performance starts to deviate from the ideal scaling as early as 16 threads (QRC on Graviton) and 32 threads (GHZ on Grace). 

To investigate the bottleneck in multithreading, we measure the $frontend\_bound$ and $backend\_bound$ metrics in GHZ on Grace using 1, 4, 8, 16, 32, and 64 threads. The backend bound increases as the number of threads increases. We observe that $frontend\_bound$ sustains or even slightly reduces from 11.4\% to eventually 9\%, while $backend\_bound$ increases from 21\% to 61\%. Furthermore, we measure $stall\_backend\_mem$ over the total CPU cycles at increased threads and the results show that the percentage of $stall\_backend\_mem$ increases from 4\% to 15.4\% as the number of threads increases. This finding again highlights the importance of matching memory subsystem with the aggregated throughput needed by all vector units like SVE or RVV on a system.

\subsection{Comparison with GPU and non-SVE CPU}
\begin{figure}
    \centering
    \subfloat[Grace]{\includegraphics[width=0.49\linewidth]{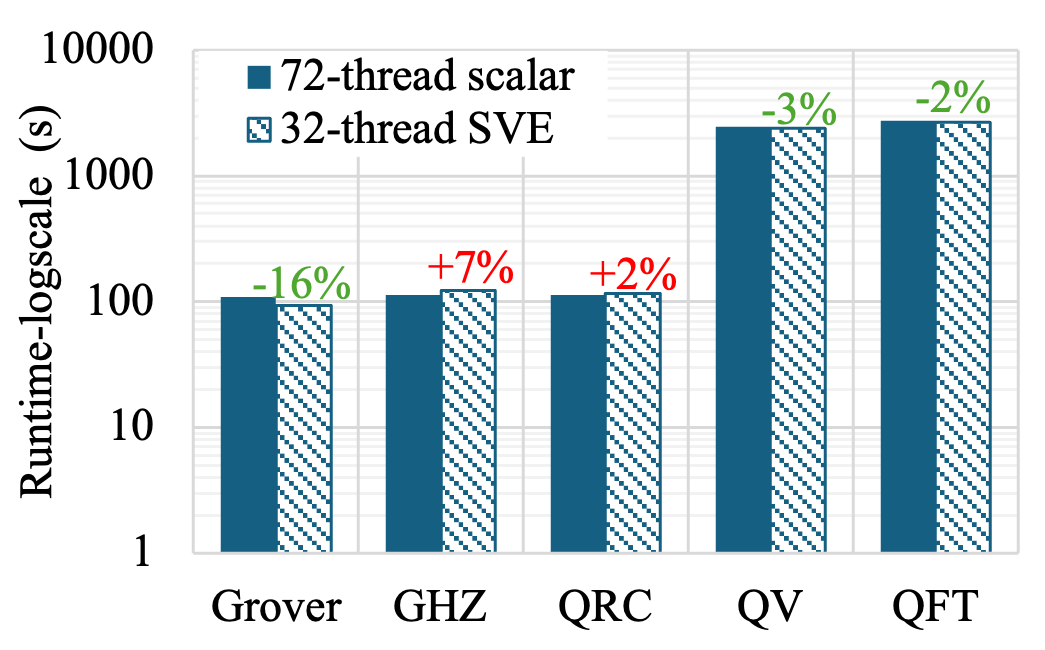}} 
    \subfloat[A64FX]{\includegraphics[width=0.49\linewidth]{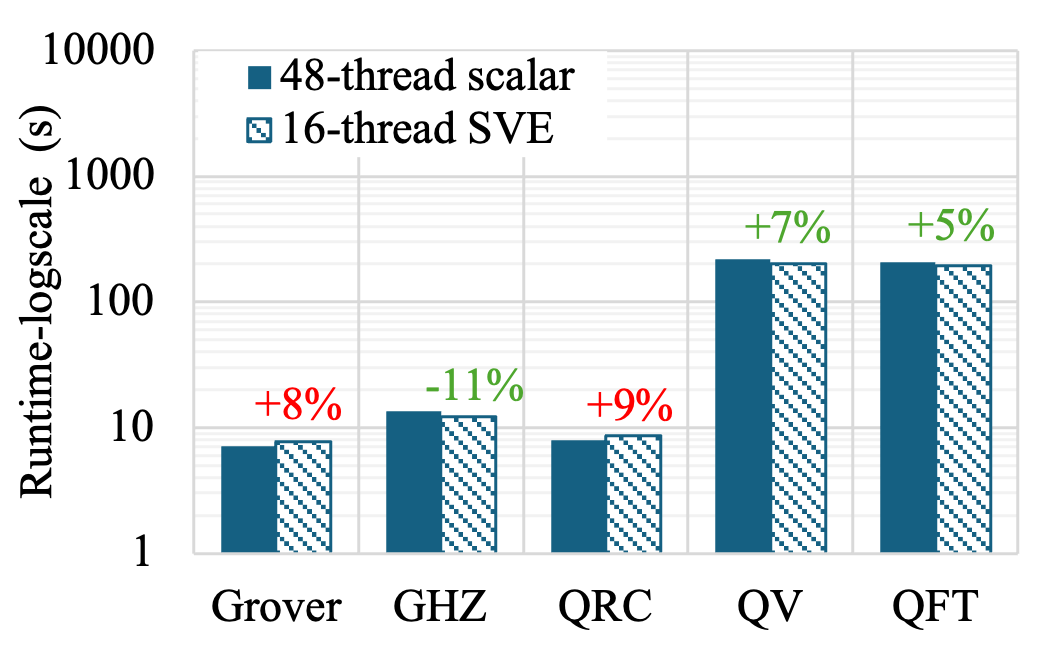}} 
    \caption{A runtime comparison of the non-vectorized version and the SVE version using $2-3\times$ fewer CPUs.} 
    \label{fig:scalar_sve}
    \vspace{-10pt}
\end{figure}

Many quantum computer simulations benefit from GPU acceleration~\cite{faj2023quantum}. Thus, our final evaluation compares SVE accelerated simulations with GPU accelerated simulations. Fig.~\ref{fig:scalar_sve} compares the computing resources needed by non-SVE CPUs and SVE-enabled CPUs to achieve similar simulation time on Grace (36-qubits circuits) and A64FX (31-qubits circuits). Without SVE, we need 72 CPUs to reach the same or even slightly lower performance than using only 32 SVE-enabled CPUs on Grace. Similar, on A64FX, we need 48 CPUs to enable the same simulations that only need 16 SVE-enabled CPUs. This $2\times$ and $3\times$ reduction of CPU cores could potentially improve system-level energy efficiency. 

We also compare the performance with the Hopper H100 GPU. Fig.~\ref{fig:GPU} shows that the SVE version is more suitable for simulating small quantum circuits. For example, Grover's algorithm with fewer than 22 qubits is constantly faster on SVE than GPU. Another advantage of using SVE comes from large capacity of CPU memory, which is needed by quantum simulators due to their exponential memory footprint. Due to limited GPU memory, the largest circuits simulated on a H100 GPU is only 31 qubits, while the SVE version supports up to 36 qubits on the CPU side.

\section{Related work}
\textbf{Application Porting and Optimizations on VLA architectures.}
ARM SVE and RISC-V Vector extension are the two main architectures that have commercially available hardware and support the VLA programming model. Langarita et al.~\cite{langarita2023porting} ported the BWA-MEM2 kernel for use in the medical field using SVE intrinsics on A64FX. Banchelli et al.~\cite{banchelli2025exploring} focus on the GEMM kernel of SeisSol and MiniFALL3D in the field of Earth Sciences on the prototype vector architecture EPAC. Browne et al.~\cite{brown2025exploring} explored Fast Fourier Transforms on the Tenstorrent Wormhole, a RISC-V accelerator with RVV. Gupta et al.~\cite{gupta2023accelerating} accelerated the im2col+GEMM kernel of CNN inference on both real hardware A64FX with SVE and also measured the impact of different vector lengths and LLC size for long vector architectures using the gem5 simulator. Blancafort et al.~\cite{blancafort2024exploiting,banchelli2025exploring} ported a CFD codes on the prototype Vitruvius VPU RVV0.7.1 and also propose a set of performance matrices for understanding vector usage. This work contributes a new VLA based design for workloads in quantum computing based on production-level Qsim and in-depth analysis on three ARM processors.  

\textbf{State-vector quantum simulators.}  
Different aspects of optimizing state-vector quantum simulators have been explored. For reduced memory usage, prior works have proposed data compression~\cite{wu2019full}, leveraging state sparsity~\cite{jaques2022leveraging}, and data format selection~\cite{fatima2021faster}, gate fusion and gate partition for avoiding separate memory traversals~\cite{fang2022efficient,quantum_ai_team_and_collaborators_2020_4023103}. mpiQulacs~\cite{tabuchi2023mpiqulacs} focuses on inter-node communication to enable distributed simulations on A64FX-based Fugaku supercomputer and they support up to 39 qubits on 512 nodes. Tsuji et al.~\cite{tsuji2022performance} use SVE as a fixed 512-bit VLS style on A64FX. 
VLS vectorization on Intel platforms has been explored extensive~\cite{smelyanskiy2016qhipster,khammassi2017qx,takahashi2023prototype}. Fatima et al.\cite{fatima2021faster} use AVX and the recursive FFT-like algorithm and optimizations on the CPU focus on cache blocking and improving memory locality. Our work focuses on VLA architecture in a production quantum simulator to gather insights for future SVE and RVV designs.


\section{Conclusion}
Vector-length agnostic architectures are emerging, as represented by ARM SVE and RISC-V vector extension. In this work, we leverage an important workload for quantum computing, quantum state-vector simulations, to understand whether high-performance portability can be achieved in a VLA design. We evaluate the current compiler support for auto-vectorization targeting ARM SVE. We then propose a VLA design, including optimization techniques such as VLEN-adaptive memory access, buffering, and fine-grained loop control. Implemented in the state-of-the-art Qsim simulator from Google, we evaluate five quantum circuits of up to 36 qubits on Grace, Graviton, and A64FX processors. By defining a set of metrics and PMU events to quantify vectorization activities, we draw insights for future VLA designs. Our VLA quantum simulator achieves up to $4.5\times$ speedup on A64FX, $2.5\times$ on Grace, and $1.5\times$ on Graviton with a single source implementation. 

\section*{Acknowledgment}
This research is supported by the European Commission under the Horizon project OpenCUBE (101092984). This work was performed under the auspices of the U.S. Department
of Energy by Lawrence Livermore National Laboratory under Contract 
DE-AC52-07NA27344 under LDRD Project 25-ERD-016. LLNL-CONF-2016001. Access to JUPITER was provided through the JUPITER Research and Early Access Program (JUREAP); JUPITER is funded by the EuroHPC Joint Undertaking, the German Federal Ministry of Research, Technology and Space, and the Ministry of Culture and Science of the German state of North-Rhine Westphalia.

\bibliographystyle{IEEEtran}
\bibliography{main}

@inproceedings{faj2023quantum,
  title={Quantum computer simulations at warp speed: Assessing the impact of gpu acceleration: A case study with ibm qiskit aer, nvidia thrust \& cuquantum},
  author={Faj, Jennifer and Peng, Ivy and Wahlgren, Jacob and Markidis, Stefano},
  booktitle={2023 IEEE 19th International Conference on e-Science (e-Science)},
  pages={1--10},
  year={2023},
  organization={IEEE}
}

@inproceedings{fatima2021faster,
  title={Faster schr{\"o}dinger-style simulation of quantum circuits},
  author={Fatima, Aneeqa and Markov, Igor L},
  booktitle={2021 IEEE International Symposium on High-Performance Computer Architecture (HPCA)},
  pages={194--207},
  year={2021},
  organization={IEEE}
}

@article{li2019quantum,
  title={Quantum supremacy circuit simulation on Sunway TaihuLight},
  author={Li, Riling and Wu, Bujiao and Ying, Mingsheng and Sun, Xiaoming and Yang, Guangwen},
  journal={IEEE Transactions on Parallel and Distributed Systems},
  volume={31},
  number={4},
  pages={805--816},
  year={2019},
  publisher={IEEE}
}

@article{smelyanskiy2016qhipster,
  title={qHiPSTER: The quantum high performance software testing environment},
  author={Smelyanskiy, Mikhail and Sawaya, Nicolas PD and Aspuru-Guzik, Al{\'a}n},
  journal={arXiv preprint arXiv:1601.07195},
  year={2016}
}

@manual{arm_sve_c_lang_ext,
  title        = {Arm C Language Extensions for SVE},
  author       = {{Arm Ltd.}},
  year         = {n.d.},
  note         = {Arm Architecture Reference Manual},
  url          = {https://developer.arm.com/documentation/101028/latest/}
}

@misc{arm-sve,
  title={Introduction to SVE Version 1.0},
  author={ARM},
  year={2024},
  howpublished={\url{https://developer.arm.com/documentation/102476/0100/Programming-with-SVE/Auto-vectorization?lang=en}}
}

@misc{armv9,
  title={Arm Architecture Reference Manual for A-profile architecture},
  author={ARM},
  year={2025},
  howpublished={\url{https://www.arm.com/architecture/cpu/a-profile/armv9}}
}

@inproceedings{stephens2016armv8,
  title={ARMv8-A next-generation vector architecture for HPC.},
  author={Stephens, Nigel},
  booktitle={Hot Chips Symposium},
  pages={1--31},
  year={2016}
}

@book{Cirq_Developers_2025, 
    title={Cirq}, 
    url={https://zenodo.org/doi/10.5281/zenodo.4062499}, 
    DOI={10.5281/ZENODO.4062499}, 
    abstractNote={Python package for writing, manipulating, and running quantum circuits on quantum computers and simulators.}, publisher={Zenodo}, author={Cirq Developers}, 
    year={2025}, 
    month=apr }

@article{arute2019quantum,
  title={Quantum supremacy using a programmable superconducting processor},
  author={Arute, Frank and Arya, Kunal and Babbush, Ryan and Bacon, Dave and Bardin, Joseph C and Barends, Rami and Biswas, Rupak and Boixo, Sergio and Brandao, Fernando GSL and Buell, David A and others},
  journal={Nature},
  volume={574},
  number={7779},
  pages={505--510},
  year={2019},
  publisher={Nature Publishing Group UK London}
}

@inproceedings{yasin2014top,
  title={A top-down method for performance analysis and counters architecture},
  author={Yasin, Ahmad},
  booktitle={2014 IEEE International Symposium on Performance Analysis of Systems and Software (ISPASS)},
  pages={35--44},
  year={2014},
  organization={IEEE}
}

@misc{qiskit2024,
      title={Quantum computing with {Q}iskit},
      author={Javadi-Abhari, Ali and Treinish, Matthew and Krsulich, Kevin and Wood, Christopher J. and Lishman, Jake and Gacon, Julien and Martiel, Simon and Nation, Paul D. and Bishop, Lev S. and Cross, Andrew W. and Johnson, Blake R. and Gambetta, Jay M.},
      year={2024},
      doi={10.48550/arXiv.2405.08810},
      eprint={2405.08810},
      archivePrefix={arXiv},
      primaryClass={quant-ph}
}

@software{quantum_ai_team_and_collaborators_2020_4023103,
  author       = {Quantum AI team and collaborators},
  title        = {qsim},
  month        = Sep,
  year         = 2020,
  publisher    = {Zenodo},
  doi          = {10.5281/zenodo.4023103},
  url          = {https://doi.org/10.5281/zenodo.4023103}
}

@article{pohl2020vectorization,
  title={Vectorization cost modeling for NEON, AVX and SVE},
  author={Pohl, Angela and Cosenza, Biagio and Juurlink, Ben},
  journal={Performance Evaluation},
  volume={140},
  pages={102106},
  year={2020},
  publisher={Elsevier}
}

@article{langarita2023porting,
  title={Porting and optimizing BWA-MEM2 using the Fujitsu A64FX processor},
  author={Langarita, Rub{\'e}n and Armejach, Adri{\`a} and Ib{\'a}{\~n}ez, Pablo and Alastruey-Bened{\'e}, Jes{\'u}s and Moret{\'o}, Miquel},
  journal={IEEE/ACM Transactions on Computational Biology and Bioinformatics},
  volume={20},
  number={5},
  pages={3139--3153},
  year={2023},
  publisher={IEEE}
}

@inproceedings{tabuchi2023mpiqulacs,
  title={mpiqulacs: A scalable distributed quantum computer simulator for arm-based clusters},
  author={Tabuchi, Akihiro and Imamura, Satoshi and Yamazaki, Masafumi and Honda, Takumi and Kasagi, Akihiko and Nakao, Hiroshi and Fukumoto, Naoto and Nakashima, Kohta},
  booktitle={2023 IEEE International Conference on Quantum Computing and Engineering (QCE)},
  volume={1},
  pages={959--969},
  year={2023},
  organization={IEEE}
}

@inproceedings{wu2019full,
  title={Full-state quantum circuit simulation by using data compression},
  author={Wu, Xin-Chuan and Di, Sheng and Dasgupta, Emma Maitreyee and Cappello, Franck and Finkel, Hal and Alexeev, Yuri and Chong, Frederic T},
  booktitle={Proceedings of the International Conference for High Performance Computing, Networking, Storage and Analysis},
  pages={1--24},
  year={2019}
}

@article{williams2009roofline,
  title={Roofline: an insightful visual performance model for multicore architectures},
  author={Williams, Samuel and Waterman, Andrew and Patterson, David},
  journal={Communications of the ACM},
  volume={52},
  number={4},
  pages={65--76},
  year={2009},
  publisher={ACM New York, NY, USA}
}

@inproceedings{blancafort2024exploiting,
  title={Exploiting long vectors with a CFD code: a co-design show case},
  author={Blancafort, Marc and Ferrer, Roger and Houzeaux, Guillaume and Garcia-Gasulla, Marta and Mantovani, Filippo},
  booktitle={2024 IEEE International Parallel and Distributed Processing Symposium (IPDPS)},
  pages={453--464},
  year={2024},
  organization={IEEE}
}

@article{banchelli2025exploring,
  title={Exploring RISC-V long vector capabilities: A case study in Earth Sciences},
  author={Banchelli, Fabio and Jurado, David and Garcia-Gasulla, Marta and Mantovani, Filippo},
  journal={Future Generation Computer Systems},
  pages={107932},
  year={2025},
  publisher={Elsevier}
}

@article{waterman2014risc,
  title={The RISC-V instruction set manual, volume I: User-level ISA, version 2.0},
  author={Waterman, Andrew and Lee, Yunsup and Patterson, David A and Asanovic, Krste},
  journal={EECS Department, University of California, Berkeley, Tech. Rep. UCB/EECS-2014-54},
  pages={4},
  year={2014}
}

@article{cross2019validating,
  title={Validating quantum computers using randomized model circuits},
  author={Cross, Andrew W and Bishop, Lev S and Sheldon, Sarah and Nation, Paul D and Gambetta, Jay M},
  journal={Physical Review A},
  volume={100},
  number={3},
  pages={032328},
  year={2019},
  publisher={APS}
}

@inproceedings{fang2022efficient,
  title={Efficient hierarchical state vector simulation of quantum circuits via acyclic graph partitioning},
  author={Fang, Bo and {\"O}zkaya, M Yusuf and Li, Ang and {\c{C}}ataly{\"u}rek, {\"U}mit V and Krishnamoorthy, Sriram},
  booktitle={2022 IEEE International Conference on Cluster Computing (CLUSTER)},
  pages={289--300},
  year={2022},
  organization={IEEE}
}

@inproceedings{khammassi2017qx,
  title={QX: A high-performance quantum computer simulation platform},
  author={Khammassi, Nader and Ashraf, Imran and Fu, Xiang and Almudever, Carmen G and Bertels, Koen},
  booktitle={Design, Automation \& Test in Europe Conference \& Exhibition (DATE), 2017},
  pages={464--469},
  year={2017},
  organization={IEEE}
}

@inproceedings{takahashi2023prototype,
  title={Prototype of a Batched Quantum Circuit Simulator for the Vector Engine},
  author={Takahashi, Keichi and Mori, Toshio and Takizawa, Hiroyuki},
  booktitle={Proceedings of the SC'23 Workshops of The International Conference on High Performance Computing, Network, Storage, and Analysis},
  pages={1499--1505},
  year={2023}
}

@article{jaques2022leveraging,
  title={Leveraging state sparsity for more efficient quantum simulations},
  author={Jaques, Samuel and H{\"a}ner, Thomas},
  journal={ACM Transactions on Quantum Computing},
  volume={3},
  number={3},
  pages={1--17},
  year={2022},
  publisher={ACM New York, NY}
}

@article{brown2025exploring,
  title={Exploring Fast Fourier Transforms on the Tenstorrent Wormhole},
  author={Brown, Nick and Davies, Jake and LeClair, Felix},
  journal={arXiv preprint arXiv:2506.15437},
  year={2025}
}

@inproceedings{gupta2023accelerating,
  title={Accelerating CNN inference on long vector architectures via co-design},
  author={Gupta, Sonia Rani and Papadopoulou, Nikela and Pericas, Miquel},
  booktitle={2023 IEEE International Parallel and Distributed Processing Symposium (IPDPS)},
  pages={145--155},
  year={2023},
  organization={IEEE}
}

@article{gomez2023hpcg,
  title={HPCG on long-vector architectures: Evaluation and optimization on NEC SX-Aurora and RISC-V},
  author={G{\'o}mez, Constantino and Mantovani, Filippo and Focht, Erich and Casas, Marc},
  journal={Future Generation Computer Systems},
  volume={143},
  pages={152--162},
  year={2023},
  publisher={Elsevier}
}

@misc{FUJITSU,
  title={FUJITSU Processor A64FX},
  howpublished={\url{https://www.fujitsu.com/global/products/computing/servers/supercomputer/a64fx/}}
}

@misc{jupiter,
  title={JUPITER Supercomputer Propels European Computing Power},
  howpublished={\url{https://www.fz-juelich.de/en/news/archive/press-release/2025/jupiter-supercomputer-propels-european-computing-power}}
}

@article{minervini2021vitruvius,
  title={Vitruvius: And area-efficient RISC-V decoupled vector accelerator for high performance computing},
  author={Minervini, Francesco and Palomar, O},
  journal={RISC-V Summit},
  year={2021}
}

@article{cavalcante2019ara,
  title={Ara: A 1-GHz+ scalable and energy-efficient RISC-V vector processor with multiprecision floating-point support in 22-nm FD-SOI},
  author={Cavalcante, Matheus and Schuiki, Fabian and Zaruba, Florian and Schaffner, Michael and Benini, Luca},
  journal={IEEE Transactions on Very Large Scale Integration (VLSI) Systems},
  volume={28},
  number={2},
  pages={530--543},
  year={2019},
  publisher={IEEE}
}

@manual{Fujitsu2020A64FX,
  title     = "A64FX Microarchitecture Manual",
  author    = {{Fujitsu Limited}},
  year      = {2020},
  edition      = {Version 1.0},
  address   = "Japan"
}

@article{coppersmith2002approximate,
  title={An approximate Fourier transform useful in quantum factoring},
  author={Coppersmith, Don},
  journal={arXiv preprint quant-ph/0201067},
  year={2002}
}

@inproceedings{grover1996fast,
  title={A fast quantum mechanical algorithm for database search},
  author={Grover, Lov K},
  booktitle={Proceedings of the twenty-eighth annual ACM symposium on Theory of computing},
  pages={212--219},
  year={1996}
}

@article{fritz2012beyond,
  title={Beyond Bell's theorem: correlation scenarios},
  author={Fritz, Tobias},
  journal={New Journal of Physics},
  volume={14},
  number={10},
  pages={103001},
  year={2012},
  publisher={IOP Publishing}
}

@inproceedings{tsuji2022performance,
  title={Performance analysis of a state vector quantum circuit simulation on A64FX processor},
  author={Tsuji, Miwako and Sato, Mitsuhisa},
  booktitle={2022 IEEE International Conference on Cluster Computing (CLUSTER)},
  pages={563--572},
  year={2022},
  organization={IEEE}
}

@inproceedings{shi2025arm,
  title={ARM SVE Unleashed: Performance and Insights Across HPC Applications on Nvidia Grace},
  author={Shi, Ruimin and Schieffer, Gabin and Gokhale, Maya and Lin, Pei-Hung and Patel, Hiren and Peng, Ivy},
  booktitle={European Conference on Parallel Processing},
  pages={33--47},
  year={2025},
  organization={Springer}
}

@inproceedings{lai2025risc,
  title={RISC-V Vectorization Coverage for HPC: A TSVC-Based Analysis},
  author={Lai, Hung-Ming and Lin, Pei-Hung and Gokhale, Maya and Peng, Ivy and Patel, Hiren and Lee, Jenq-Kuen},
  booktitle={Proceedings of the SC'25 Workshops of the International Conference for High Performance Computing, Networking, Storage and Analysis},
  pages={1676--1683},
  year={2025}
}





\twocolumn[%
{\begin{center}
\Huge
Appendix: Artifact Description/Artifact Evaluation        
\end{center}}
]


\appendixAD

\section{Overview of Contributions and Artifacts}

\subsection{Paper's Main Contributions}

\artexpl{
Provide a list of all main contributions of the paper.
}

\begin{description}
\item[$C_1$] We identify inefficiencies in the current compiler support for VLA auto-vectorization in quantum simulations.
\item[$C_2$] We propose a VLA design for quantum state-vector simulations and optimization techniques, including VLEN-adaptive memory access, buffering, and fine-grained loop control.
\item[$C_3$] We provide an implementation in Google’s Qsim, and evaluate five quantum circuits of up to 36 qubits across ARM processors, achieving up to 4.5 $\times$ speedup on A64FX, 2.5 $\times$ on Nvidia Grace, and 1.5 $\times$ on Graviton and high scalability up to 288 threads on the JUPITER supercomputer node.
\item[$C_4$] We define a set of metrics and PMU events to quantify vectorization activities and provide insights for future VLA designs.
\end{description}

\subsection{Computational Artifacts}

\artexpl{
List the computational artifacts related to this paper along with their respective DOIs. Note that all computational artifacts may be archived under a single DOI.
}

\begin{description}
\item[$A_1$] https://zenodo.org/records/18661525
\end{description}

\artexpl{
Provide a table with the relevant computational artifacts, 
highlight their relation to the contributions (from above) and 
point to the elements in the paper that are reproducible by each artifact, e.g., 
which figures or tables were generated with the artifact.
}

\begin{center}
\begin{tabular}{rll}
\toprule
Artifact ID  &  Contributions &  Related \\
             &  Supported     &  Paper Elements \\
\midrule
$A_1$   &  $C_1$ $C_2$ $C_3$ $C_4$ & Table 4 \\
        &        & Figures 2, 6-15 \\

\bottomrule
\end{tabular}
\end{center}

\section{Artifact Identification}

\artexpl{
Provide the following six subsections for each computational artifact $A_i$.
}

\newartifact

\artrel

\artexpl{
    Briefly explain the relationship between the artifact and contributions.
}

The artifact A1 is the main source code of this work. By setting backend choice, it can evaluate the capability of auto-vectorization by the compiler, supporting the contribution C1. It provides the implementation of SVE intrinsics version with VLA-specific optimizations, supporting C2 and C3. The
lightweight libperf library is designed to support C4.

\artexp

\artexpl{
Provide a higher-level description of what outcome to expect from the corresponding experiments. Provide an explanation of how the results substantiate the main contributions.
}
The runtimes over five circuits are similar in the scenarios with/without auto-vectorization. Compared to the runtime obtained by auto-vectorization, the optimized time by SVE QSim reaches 1.5-4.5×, depending on different SVE vector lengths. Counting PMU
events, the higher proportion of vector instructions and vector
activites can be observed.

\arttime

\artexpl{
Estimate the time required to reproduce the artifact, providing separate estimates for the individual steps: Artifact Setup, Artifact Execution, and Artifact Analysis.
}

The longest expected computational time traversing five circuits in 32 qubits per time is around 300 seconds in SVE QSim and 1200 seconds in Navie QSim. The time requires doubles with each additional qubit.
   
\artin
\label{A1:setup}
\artinpart{Hardware}

\artexpl{
Specify the hardware requirements and dependencies (e.g., a specific interconnect or GPU type is required).
}
Our implementation can be performed on ARM CPU equipped with SVE. The memory capability requires more than $2^{(N+3)}$ B for $N$-qubit simulation. The experiments were performed on Nvidia Grace, Graviton and A64FX CPU.
\artinpart{Software}

\artexpl{
Introduce all required software packages, including the computational artifact. For each software package, specify the version and provide the URL.
}
Python 3.9+ with Cirq, Numpy, pybind11, and setuptools package; cmake 3.26.5; GNU compiler (GCC12 and above) is used for experiments in this paper. To enable perf event, perf v5.14.0 is needed and \texttt{perf\_event\_paranoid} value of less than 1.


\artexpl{
Describe the datasets required by the artifact. Indicate whether the datasets can be generated, including instructions, or if they are available for download, providing the corresponding URL.
}
\artinpart{Installation and Deployment} Build the project by \texttt{pip install -e .}
\artexpl{
Detail the requirements for compiling, deploying, and executing the experiments, including necessary compilers and their versions.
}

\artcomp

\artexpl{
Provide an abstract description of the experiment workflow of the artifact. It is important to identify the main tasks (processes) and how they depend on each other. 

A workflow may consist of three tasks: $T_1, T_2$, and $T_3$. The task $T_1$ may generate a specific dataset. This dataset is then used as input by a computational task $T_2$, and the output of $T_2$ is processed by another task $T_3$, which produces the final results (e.g., plots, tables, etc.). State the individual tasks $T_i$ and provide their dependencies, e.g., $T_1 \rightarrow T_2 \rightarrow T_3$.

Provide details on the experimental parameters. How and why were parameters set to a specific value (if relevant for the reproduction of an artifact), e.g., size of dataset, number of data points, input sizes, etc. Additionally, include details on statistical parameters, like the number of repetitions.
}
To obtain the comparison results of non/auto-vectorized/, set \texttt{CMAKE\_CXX\_FLAGS} to one of: NonVec, SVE, Neon options; Recompile and navigate to \texttt{benchmark} directory, executing \texttt{INSTR=4 run\_qsim\_leq32.py} with the experimental parameter 31 qubits, fused 3. For SVE QSim performance, \texttt{INSTR=3} switches to SVE intrinsics backend \texttt{./benchmark/run\_multibench.sh} specifies the the
experimental parameters running on Grace CPU and similar execution steps for other ARM CPUs when adequate memory and threads. \texttt{run\_qsim\_GPU.py} can execute quantum simulation with CUDA.

\artout
Execution time and PMU event counts will be printed out and can be saved to a log file. \texttt{./benchmark/parse\_output.sh *.log} can extract PMU event counts and calculate the average. The results in the paper represent the average of five runs.

\newpage
\appendixAE

\arteval{1}
\artin

\artexpl{
Provide instructions for installing and compiling libraries and code. 
Offer guidelines on deploying the code to resources.
}

Please download and uncompress the source code,
\begin{lstlisting}[language=bash]
wget "https://zenodo.org/records/18661525/files/qsim-arm-sve-intrinsics.zip?download=1" -O qsim-arm-sve-intrinsics.zip
unzip qsim-arm-sve-intrinsics.zip 
cd qsim-arm-sve-intrinsics
\end{lstlisting}
Please create the python environment and install qsim and  its dependency 
\begin{lstlisting}[language=bash]
python -m venv qsim
source qsim/bin/activate
cd libperf/ && make && cd ..
pip install -e .
\end{lstlisting}

Change \texttt{./pybind\_interface/basic/CMakeLists.txt} Line 7 \texttt{CMAKE\_CXX\_FLAGS} to non/auto-vectorized compiler flags,
\begin{itemize}
    \item Non-vectorized: \texttt{-O3 -fno-tree-vectorize -fno-tree-loop-vectorize -fno-tree-slp-vectorize}
    \item Auto-vectorized(default): \texttt{-march=armv8.5-a+sve -mcpu=neoverse-v2}
\end{itemize}
\artcomp
The format to execute the simulator,
\begin{lstlisting}[language=bash]
INSTR=<3/4> python benchmark/<script>.py -b <circuit> -q <qubits> -t <threads> -f <2/3/4/5> -r <iterations>
\end{lstlisting}
The usage of provided scripts,
\begin{itemize}
    \item \texttt{run\_qsim\_leq32} is running a circuit with less than or equal to 32 qubits.
    \item \texttt{run\_qsim\_gt32.py} is running a circuit with greater than 32 qubits.
    \item \texttt{correctness.py} is validation of the results.
    \item \texttt{run\_qsim\_GPU.py} is running a circuit in signle GPU.
\end{itemize}

\texttt{INSTR=3} runs QSIM SVE; \texttt{INSTR=4} runs the naive version of Qsim by non/auto-vectorized compiling.
\texttt{-b <circuit>} has the option: grover, ghz, QRC, qft, qv.

\artexpl{
Describe the experiment workflow. 
If encapsulated within a workflow description or equivalent (such as a makefile or script), clearly outline the primary tasks and their interdependencies. Detail the main steps in the workflow. Merely instructing to “Run script.sh” is inadequate.
}


\artexpl{
\begin{itemize}
    \item Provide a description of the expected results and a methodology for evaluating these results. 
    \item Explain how the expected results from the experiment workflow correlate with the contributions stated in the article. 
    \item For example, if the article presents results in a figure, the artifact evaluation should also produce a similar figure, depicting the same generalizable outcome. Authors must focus on these aspects to reduce the time required for others to understand and verify an artifact.
\end{itemize}
}


\end{document}